\documentclass[aps,pra,twocolumn,showpacs,superscriptaddress]{revtex4} 
 
\usepackage{graphicx} 
\usepackage{subfigure} 
\usepackage{braket} 
\usepackage{amsmath} 
\usepackage{amssymb} 
\usepackage{bm} 
 
\begin{document} 
 
\title{Interaction of intense vuv radiation with large xenon clusters} 
 
\author{Zachary~B.~Walters} 
\affiliation{Department of Physics and JILA, University of 
Colorado, Boulder, Colorado 80309-0440, USA} 
\author{Robin~Santra} 
\affiliation{Argonne National Laboratory, Argonne, Illinois 60439, USA} 
\author{Chris~H.~Greene} 
\affiliation{Department of Physics and JILA, University of 
Colorado, Boulder, Colorado 80309-0440, USA} 
 
\date{\today} 
 
\begin{abstract} 
The interaction of atomic clusters with short, intense pulses of laser 
light to form extremely hot, dense plasmas has attracted extensive 
experimental and theoretical interest.  The high density of atoms 
within the cluster greatly enhances the atom--laser interaction, while 
the finite size of the cluster prevents energy from escaping the 
interaction region. Recent technological advances have allowed 
experiments to probe the laser--cluster interaction at very high photon 
energies, with interactions much stronger than suggested by theories 
for lower photon energies.  We present a model of the laser--cluster  
interaction which  uses non-perturbative R-matrix techniques to calculate 
inverse bremsstrahlung and photoionization cross sections for 
Herman-Skillman atomic potentials.  We describe the evolution of the 
cluster under the influence of the processes of inverse bremsstrahlung 
heating, photoionization, collisional ionization and recombination, and 
expansion of the cluster.  We compare charge state distribution, 
charge state ejection energies, and total energy absorbed with the Hamburg 
experiment of Wabnitz {\em et al.} [Nature {\bf 420}, 482 (2002)] and
ejected  
electron spectra with Laarmann {\em et al.} [Phys. Rev. Lett. {\bf
95}, 063402 (2005)].  
 
\end{abstract} 
 
 
\maketitle 
 
\section{Introduction} 
 
At sufficiently low temperatures and sufficiently high density, atoms
and molecules  
in the gas phase begin to form bound systems, or clusters
\cite{Scie96,SuKo98}.  
Some clusters consist of only a few monomers; others contain many
millions of atoms or   
molecules. In this sense, clusters, which are typical nanomaterials,
represent a natural   
link between simple atoms and condensed matter. However, they do not
simply mimic the   
properties of their constituents nor of the bulk they converge
to. Clusters are unique.   
This is highlighted, for example, by their interaction with intense
electromagnetic radiation.  
 
The majority of corresponding experiments were carried out using laser pulses 
in the near-infrared, with photon energies of about $1$~eV and pulse
durations on the  
order of $100$ fs \cite{HuDi98}. At pulse intensities of
$10^{16}$~W/cm$^{2}$ or higher,  
noble gas clusters consisting of krypton or argon atoms absorb the laser pulse 
energy extremely efficiently. They are turned into nanoplasmas
accompanied by high  
ionic charge states and strong x-ray emission \cite{DiDo95}. In xenon
clusters, the  
production of extremely hot (keV) electrons was observed
\cite{ShDi96}. The hot 
nanoplasmas undergo complete fragmentation. 
Experimental studies of the associated dynamics have been carried out  
for Ar and Xe clusters by Lezius {\em et al.} \cite{LeDo98}: ions
with kinetic energies  of up to $1$~MeV are found to be ejected from the expanding
clusters.   
 
The high density of atoms in the cluster greatly enhances the atom--laser 
interaction over that of lone atoms, while the finite size of the 
clusters ensures that energy absorbed by the cluster is largely 
constrained to stay within the interaction region, not carried 
off by a large heat bath, as occurs with materials in bulk. 
These properties in combination allow the laser--cluster interaction to 
create extremely hot, dense plasmas, which may in turn serve as 
sources for high-energy particles or photons. 
 
Several groups performed extensive numerical simulations with the
purpose of identifying  
the relevant heating mechanisms. Ditmire {\em et al.} \cite{DiDo96}
pointed out that   
collisional heating dominates. More precisely, the ionized electrons
inside the cluster are  
quasifree but can absorb photons whenever they are being scattered by
ions. This process  
is referred to as {\em inverse bremsstrahlung}
\cite{SeHa73}. Inelastic electron--ion  
collisions of the ($e,2e$) type contribute to the high ionic charge
states observed. Other  
authors \cite{RoSc97,LaJo99,SiRo02} concentrated on the interplay
between the strong  
quasistatic electric field of the laser and the Coulomb field of
neighboring ions.  
This interplay can lead to {\em enhanced ionization}, first
discovered in diatomic  
molecules \cite{SeIv95,ZuBa95}. The relative importance of enhanced
ionization is somewhat  
difficult to assess, since in Refs.~\cite{RoSc97,LaJo99,SiRo02}
collisional heating was  
not considered and no comparison was provided between the simulations
and available  experimental data. 
 
Little is known about laser--cluster interactions at uv or higher
photon energies. The  
destructive impact of laser pulses with a peak intensity of almost
$10^{19}$~W/cm$^{2}$  
at a wavelength of $248$~nm was demonstrated by McPherson {\em et al.}
\cite{McTh94}.  
However, intense laser fields at even higher photon energies have not
been accessible until  
very recently. In 2000, the first lasing---in a free-electron laser (FEL)---at 
$\lambda = 109$~nm was reported \cite{AnAu00}. The FEL is part of the
TESLA Test Facility  
(TTF) in Hamburg, Germany. (One of the major objectives of the TTF is
the development of the  
technology for an ultrabright x-ray laser.) The new vuv laser source
has already displayed  
its capability for exploring interesting physics: Motivated by the
outstanding properties  
of the radiation generated by the TTF FEL, documented in
Ref.~\cite{AyBa02}, experiments   
were performed \cite{wabnitz02a} in which Van der Waals clusters of
xenon atoms were exposed   
to 12.7-eV vuv photons, an energy range which had been previously unexplored. 
 
Each pulse in the experiment lasted about $100$~fs. The highest
intensity in the experiment   
was about $7 \times 10^{13}$~W/cm$^{2}$.  Under these conditions, isolated Xe  
atoms were found to produce only singly charged ions (see
Refs.~\cite{WaCa05} and \cite{SaGr04}  
for recent developments). In large clusters, however, each atom was
found to absorb up to   
$400$~eV, corresponding to 30 photons, and charge states of up to 8
plus were detected.  
 
These results were very surprising. The dominant processes in most 
models of infrared laser--cluster interactions are field ionization of 
atoms by the strong electric field of the laser, and heating of the 
cluster through inelastic dephasing collisions by electrons oscillating  
in the laser electric field. Both of these processes are strongly inhibited 
by the high frequency of the vuv photons. A relevant quantity in this context 
is the ponderomotive potential \cite{burnett89} of an electron
oscillating in a   
laser field. At a laser intensity of $7 \times 10^{13}$~W/cm$^2$, the 
ponderomotive potential is only $62$~meV, which is smaller than the 
ionization potential of atomic xenon (12.1 eV \cite{BrVe01}) by three orders  
of magnitude. 
 
In a previous Letter, two of us \cite{santra} proposed that the
additional heating   
was due to the effects of atomic structure on the inverse bremsstrahlung 
rates.  However, that initial study calculated the inverse bremsstrahlung 
rates using perturbation theory for both the electron--photon interaction 
and the electron--ion interaction.  The method implemented for this paper 
instead treats the electron--ion interaction nonperturbatively using 
variational R-matrix methods, while retaining first-order perturbation 
theory for the electron--photon interaction.  The resulting photoionization 
and inverse bremsstrahlung cross sections calculated are expected to be 
more realistic, even though our description remains at the 
independent electron level. 
 
In this paper, we present a model of the laser--cluster interaction 
that takes atomic structure and the effects of plasma screening into 
account more fully than previous approaches to the subject. Considering the 
limitations of the model and the poor experimental characterization of the  
FEL radiation \cite{Moeller}, we achieve good agreement with the
Hamburg results.    
We track photoionization, collisional ionization and recombination, inverse  
bremsstrahlung heating, evaporation of 
electrons from the cluster, and expansion of the cluster due to 
hydrodynamic pressure of hot electrons and Coulomb repulsion  through 
the duration of the laser pulse and, later, as the cluster undergoes 
a Coulomb explosion. 
 
Atomic units are used throughout, unless otherwise noted. 
 
\section{Photoionization}\label{sec:Photoionization} 
 
A novel feature of the Hamburg experiment is that the 12.7-eV photons 
are sufficient to overcome the 12.1-eV ionization potential of neutral 
xenon.  Therefore, although the oscillating electric field is too weak 
for the xenon atoms to undergo field ionization, as occurs in the infrared
domain, there is still an efficient optical process for creating 
Xe$^+$. 
 
Friedrich \cite{friedrich} gives the cross section for the transition
from the bound state  
$\Ket{\phi_i}$ to the continuum state $\Ket{\phi_f}$ as 
\begin{equation} 
\sigma_{fi}(E)=4\pi^2 \alpha \omega 
\vert \vec{\pi} \cdot \vec{r}_{fi} \vert ^2 \; , 
\label{eq:photoionization} 
\end{equation} 
where $\alpha$ is the fine-structure constant, $\omega$ is the photon energy, 
$\vec{\pi}$ is the polarization vector for the radiation, and 
$\vec{r}_{fi}=\Bra{\phi_{f}} \vec{r}\Ket{\phi_i}$ is the 
dipole matrix element coupling the initial and final states of the electron. 
The wave function of the photoelectron in Eq.~(\ref{eq:photoionization}) 
is energy-normalized. $E$ stands for the kinetic energy of the photoelectron. 
 
For linearly polarized light, chosen without loss of generality to be 
polarized in the $\hat{z}$ direction, the matrix element that must be 
found is 
\begin{equation} 
\Bra{\phi_{f}}z\Ket{\phi_{i}}= I_{R}(l_{i},l_{f}) \int d \Omega 
Y^{\ast}_{l_f m_f}(\Omega) \cos{(\theta)} 
Y_{l_i m_i}(\Omega) \; . 
\label{PI1a} 
\end{equation} 
Here, 
\begin{equation} 
\label{PI1b} 
I_{R}(l_{i},l_{f}) = \int_{0}^{\infty} d r U_{f}(r) r  U_i(r) \; , 
\end{equation} 
where $U(r)=r R(r)$ denotes the rescaled radial wave function. 
Equations~(\ref{eq:photoionization}) and (\ref{PI1a}) refer to
specific angular momentum   
quantum numbers $l$ and $m$ for the initial and final states. At a
photon energy of 12.7 eV,  
only the 5p electrons of xenon can respond to the radiation
field. Hence, we can focus on  
a subshell with fixed $l_i$ (here, $l_i = 1$). Let $q$ stand for the
number of electrons in   
this subshell. Then, within the independent particle model, after
averaging over the initial   
and summing over the final one-electron states, the total atomic
photoionization cross section is given by  
\begin{eqnarray} 
\label{PICrossSection} 
\sigma_{\text{PI}} & = & q \frac{4}{3}\pi^2 
\frac{\alpha \omega}{2 l_i +1} \\ 
& & \times  
\left\{l_{i} I^2_{R}(l_{i},l_{i}-1) + (l_{i}+1) I^2_{R}(l_{i},l_{i}+1)\right\} \; . 
\nonumber 
\end{eqnarray} 
The identities \cite{rotenberg,edmonds} 
\begin{widetext} 
\begin{equation} 
\label{rotenberg0} 
\int d\Omega Y^{\ast}_{l_1 m_1}(\Omega) \cos{(\theta)} Y_{l_2 m_2}(\Omega)  
= \sqrt{(2 l_1 + 1)(2 l_2 + 1)} (-1)^{-m_1} \begin{pmatrix}l_1 & 1 & l_2 \\ -m_1 & 0 & m_2 \end{pmatrix} 
\begin{pmatrix}l_1 & 1 & l_2 \\ 0 & 0 & 0 \end{pmatrix} \; ,  
\end{equation} 
\begin{equation} 
\sum_{m_1,m_2} 
\begin{pmatrix}l_1 & l_2 & l_3 \\ m_1 & m_2 & m_3\end{pmatrix} 
\begin{pmatrix}l_1 & l_2 & l^{\prime}_3 \\ m_1 & m_2 & 
m^{\prime}_{3}\end{pmatrix} 
=\frac{\delta(l_3,l^{\prime}_3) \delta(m_3,m^{\prime}_3)}{2 l_3 + 1} \; ,  
\label{rotenberg1} 
\end{equation} 
\end{widetext} 
and 
\begin{equation} 
(2 l_1 +1)(2 l_2 + 1) 
\begin{pmatrix}l_1 & 1 & l_2 \\ 0 & 0 & 0\end{pmatrix}^2 
= 
\begin{cases} 
l_1+1 & \text{if $l_2=l_1+1$} \\ 
l_1 & \text{if $l_2=l_1 - 1$} \\ 
0 & \text{otherwise} \\ 
\end{cases} 
\label{threejsum} 
\end{equation} 
have been exploited in the derivation of Eq.~(\ref{PICrossSection}). 
$\begin{pmatrix}l_1 & l_2 & l_3 \\ m_1 & m_2 & m_3\end{pmatrix}$ 
represents the Wigner 3-j symbol, related to the Clebsch-Gordan coefficient by 
\begin{eqnarray} 
\begin{pmatrix}l_1 & l_2 & l \\ m_1 & m_2 & m\end{pmatrix} 
& = & \frac{(-1)^{l_1-l_2-m}}{\sqrt{2 l +1}} \\ 
& & \times <l_{1} m_{1} l_{2} m_{2} | l_{1} l_{2} l -m> \; . \nonumber 
\end{eqnarray} 
 
The radial integrals $I_{R}(l_{i},l_{f})$ [Eq.~(\ref{PI1b})] were calculated  
in the acceleration representation, using wave functions generated in
a variational  
eigenchannel R-matrix calculation \cite{aymar}, using a 
B-spline basis set to describe electrons that experience a
Hartree-Slater atomic potential.   
This potential, which was calculated employing the program by Herman
and Skillman \cite{HeSk63},  
is spherically symmetric. Its eigenstates may therefore be chosen as
eigenstates of orbital angular momentum. 
 
As a consequence of efficient photoionization, the electrons and ions inside  
the cluster form a dense, nanoscale-size plasma already at an early
stage of the  
laser pulse. This plasma has the effect of screening the atomic
potential from both bound and continuum electrons.  This lowers the
ionization potential and changes   
both the initial- and final-state electron wave functions. Because of
this, cross sections 
for photoionization become larger as the screening length in the plasma 
becomes shorter.  With sufficient screening, it becomes possible for 
ions to undergo additional photoionization. 
 
To account for this process, the screened radial matrix elements were 
calculated using the same R-matrix methods as for isolated Xe atoms. 
However, before the initial- and final-state wave functions were 
calculated, the Herman-Skillman potential was multiplied by a Debye 
screening factor $\exp{(-r/\lambda_D)}$. (The electron Debye length is defined 
as $\lambda_D = \sqrt{T/(4 \pi n_e)}$ \cite{Krue03}. The electron 
temperature $T$ in this expression is measured in units of 
energy. $n_e$ is the electron 
density.) Both the matrix elements and the corresponding 
photoionization potentials were  
then spline-interpolated in the process of calculating the 
photoionization cross section at a given screening length.  For most 
of this paper, this screening length was restricted to be no less than 
the Wigner-Seitz radius of xenon at liquid density, 4.64 bohr. A
discussion of shorter 
screening lengths is given in a later section.  
 
\section{Inverse Bremsstrahlung Heating} 
 
A second effect of having a high density of free electrons in the 
cluster plasma is that these electrons can themselves undergo 
both stimulated and inverse bremsstrahlung, creating a second 
mechanism through which laser energy can be deposited into the 
cluster. Stimulated (inverse) bremsstrahlung refers to photon 
emission into (absorption from) the laser mode by an electron  
colliding with an ion in the plasma.  
 
We treat the collisions of an electron with the cluster ions  
as independent events in both time and space. This allows us to  
focus on a single collision of an electron with a single ion embedded  
in the plasma. The cross section per unit energy for a free-free transition  
from initial state $\ket{\phi_{E',l',m'}}$ to final state $\ket{\phi_{E,l,m}}$ 
can be shown, using Fermi's golden rule, to equal 
\begin{equation} 
\label{free1} 
\sigma_{E,l,m \leftarrow E',l',m'}=\frac{4 \pi^2 \alpha}{\omega^{3}}  
\left|\bra{\phi_{E,l,m}}\frac{\partial V}{\partial z}\ket{\phi_{E',l',m'}}\right|^2 \; .  
\end{equation} 
In the case of photon emission (absorption), 
$E = E' - \omega$ ($E = E' + \omega$). 
Equation~(\ref{free1}) describes the interaction of linearly 
polarized radiation in the acceleration representation. 
$V$ is the plasma-screened atomic potential experienced by the  
scattered electron. 
 
As with photoionization cross sections, radial wave functions were 
calculated using a nonperturbative eigenchannel R-matrix approach. 
Matrix elements between the energy-normalized wave functions were then 
calculated in the acceleration gauge, where the $1/r^2$ 
long-range dependence of $\partial V/\partial z$ ensures that 
the radial integral will converge, although the continuum electron 
wave functions are not spatially normalizable. 
 
Although microscopic reversibility ensures that absorption and emission 
cross sections coincide, stimulated free-free transitions act as a powerful  
heating process because lower energy states are more highly populated than higher 
energy states in a thermal distribution. Heating rates can then be 
calculated for any given electron distribution. In this study, we assume that 
after each photon absorption or emission event the electron gas reequilibrates 
rapidly as a consequence of frequent electron--electron collisions. Thus, 
the electron probability distribution $\rho(E)$ may be written at all times 
during the laser pulse as a Maxwell-Boltzmann distribution: 
\begin{equation} 
\label{MB1} 
\rho(E) = 2 \sqrt{\frac{E}{\pi T^3}} e^{-E/T} \; . 
\end{equation} 
 
The cross section defined in Eq.~(\ref{free1}) describes a free-free
transition  
between orbital angular momentum eigenstates. We therefore introduce
$\rho(E,l,m)$,  
which is the probability per unit energy to find an electron in the state 
$\ket{\phi_{E,l,m}}$. Clearly, $\rho(E) = \sum_{l,m} \rho(E,l,m)$. If the 
wave function is normalized within a large sphere of radius $R$ (not to be 
confused with the cluster radius), then the largest $l$ that contributes 
to this sum at a given kinetic energy $E$ is $l_{max} = R\sqrt{2 E}$
\cite{comment1}.  
Since, in thermal equilibrium, $\rho(E,l,m)$ can depend only on $E$,
we see that  
\begin{equation} 
\label{MB2} 
\rho(E,l,m) = \frac{\rho(E)}{2 R^2 E}  
\end{equation} 
in the limit of large $R$ ($l_{max} \gg 1$). 
 
We are interested in radiation-induced heating, i.e. in the change of
the electron  
temperature due to photon absorption and emission. To this end, we
will derive from  
\begin{equation} 
\label{RET1} 
\frac{\partial T}{\partial t} = \frac{2}{3} \int_{0}^{\infty} dE E \sum_{l,m} 
\frac{\partial \rho(E,l,m)}{\partial t} 
\end{equation} 
a rate equation for the electron temperature, expressed in terms of
the cross sections  
for stimulated and inverse bremsstrahlung [Eq.~(\ref{free1})]. When
writing down the  
equation for the time evolution of $\rho(E,l,m)$, we must take into
consideration  
that $\rho(E,l,m)$ refers to (spherical) box normalization, while the
cross section  
per unit energy in Eq.~(\ref{free1}) is based on energy-normalized
wave functions.  
For the sake of consistency, it is necessary to change the initial
state in the  
free-free radiative transition in Eq.~(\ref{free1}) from energy
normalization to   
box normalization. This has the effect of multiplying the cross section by  
$\pi\sqrt{2 E}/R$.  
 
Hence, in the presence of $N_a$ atomic scatterers (within the
normalization volume)  
and a laser beam of intensity $I$, the rate of change of $\rho(E,l,m)$
is given by  
\begin{widetext} 
\begin{eqnarray} 
\frac{\partial \rho(E,l,m)}{\partial t} & = & N_a \frac{I}{\omega} \frac{\sqrt{2}\pi}{R}  
\sum_{l',m'} \left\{ \sigma_{E,l,m \leftarrow E-\omega,l',m'} \sqrt{E-\omega} \rho(E-\omega,l',m')  
+ \sigma_{E,l,m \leftarrow E+\omega,l',m'} \sqrt{E+\omega} \rho(E+\omega,l',m') 
\right. \nonumber \\ 
& & - \left. \sigma_{E+\omega,l',m' \leftarrow E,l,m} \sqrt{E} \rho(E,l,m) 
- \sigma_{E-\omega,l',m' \leftarrow E,l,m} \sqrt{E} \rho(E,l,m) \right\} \; . 
\label{KEQ1} 
\end{eqnarray} 
\end{widetext} 
The first row in the curly brackets in Eq.~(\ref{KEQ1}) describes the population of  
$\ket{\phi_{E,l,m}}$ via photoabsorption (photoemission) from states with energy 
$E-\omega$ ($E+\omega$); the second row describes the depopulation of 
$\ket{\phi_{E,l,m}}$ due to photoabsorption and photoemission from this state.  
Equation~(\ref{KEQ1}) implies a nondegenerate electron gas. 
 
An electron state with energy $E$, then, communicates with states of 
energy $E-\omega$, which are on average more densely populated 
than itself. Since the absorption- and emission cross sections are 
equal, this tends to populate the state of energy $E$ and depopulate 
the states of energy $E-\omega$, resulting in a net heating 
process. The state will also communicate with states of energy $E+\omega$,  
which are less densely populated than itself, thereby again 
tending to populate the higher-energy states while depopulating the 
lower-energy state.  
 
Combining Eqs.~(\ref{MB1}), (\ref{MB2}), (\ref{RET1}), and (\ref{KEQ1}), 
we are led in a natural way to the following definition of the inverse 
bremsstrahlung cross section (per unit energy): 
\begin{equation} 
\label{IBS1} 
\sigma_{E+\omega \leftarrow E} = \sum_{l,m} \sum_{l',m'} \sigma_{E+\omega,l,m \leftarrow E,l',m'} \; . 
\end{equation} 
Using Eqs.~(\ref{rotenberg0}), (\ref{rotenberg1}), (\ref{threejsum}), and (\ref{free1}), 
this can be written as 
\begin{eqnarray} 
\label{IBS2} 
\sigma_{E+\omega \leftarrow E} & = & \frac{4}{3}\pi^2 \frac{\alpha}{\omega^3} \\ 
& & \times \sum_{l} 
\left\{l J^2_{R}(l,l-1) + (l+1) J^2_{R}(l,l+1)\right\} \; , 
\nonumber 
\end{eqnarray} 
where  
\begin{equation} 
J_{R}(l,l') = \int_{0}^{\infty} d r U_{E+\omega,l}(r) \frac{d V}{d r} U_{E,l'}(r) \; . 
\end{equation} 
The rate of change of the electron temperature due to inverse bremsstrahlung is then 
\begin{eqnarray} 
\frac{\partial T}{\partial t} & = & \frac{2}{9} n_a I \left(\frac{2\pi}{T}\right)^{3/2} 
\left[1-e^{-\omega/T}\right] \nonumber \\ 
& & \times \int_{0}^{\infty} dE e^{-E/T} \sigma_{E+\omega \leftarrow E} \; . 
\label{IBS3} 
\end{eqnarray} 
The parameter $n_a$ stands for the number of atoms per unit volume. In general, the ions 
in the plasma are not all in the same charge state. Denoting the fraction of Xe$^{i+}$ by  
$f^{(i)}$, $\sigma_{E+\omega \leftarrow E}$ in Eq.~(\ref{IBS3}) is replaced with 
$\sum_{i} f^{(i)} \sigma^{(i)}_{E+\omega \leftarrow E}$, where $\sigma^{(i)}_{E+\omega \leftarrow E}$ 
is the inverse bremsstrahlung cross section in the field of Xe$^{i+}$. 
 
\begin{figure} 
	\begin{center} 
	\includegraphics[width=3.375in]{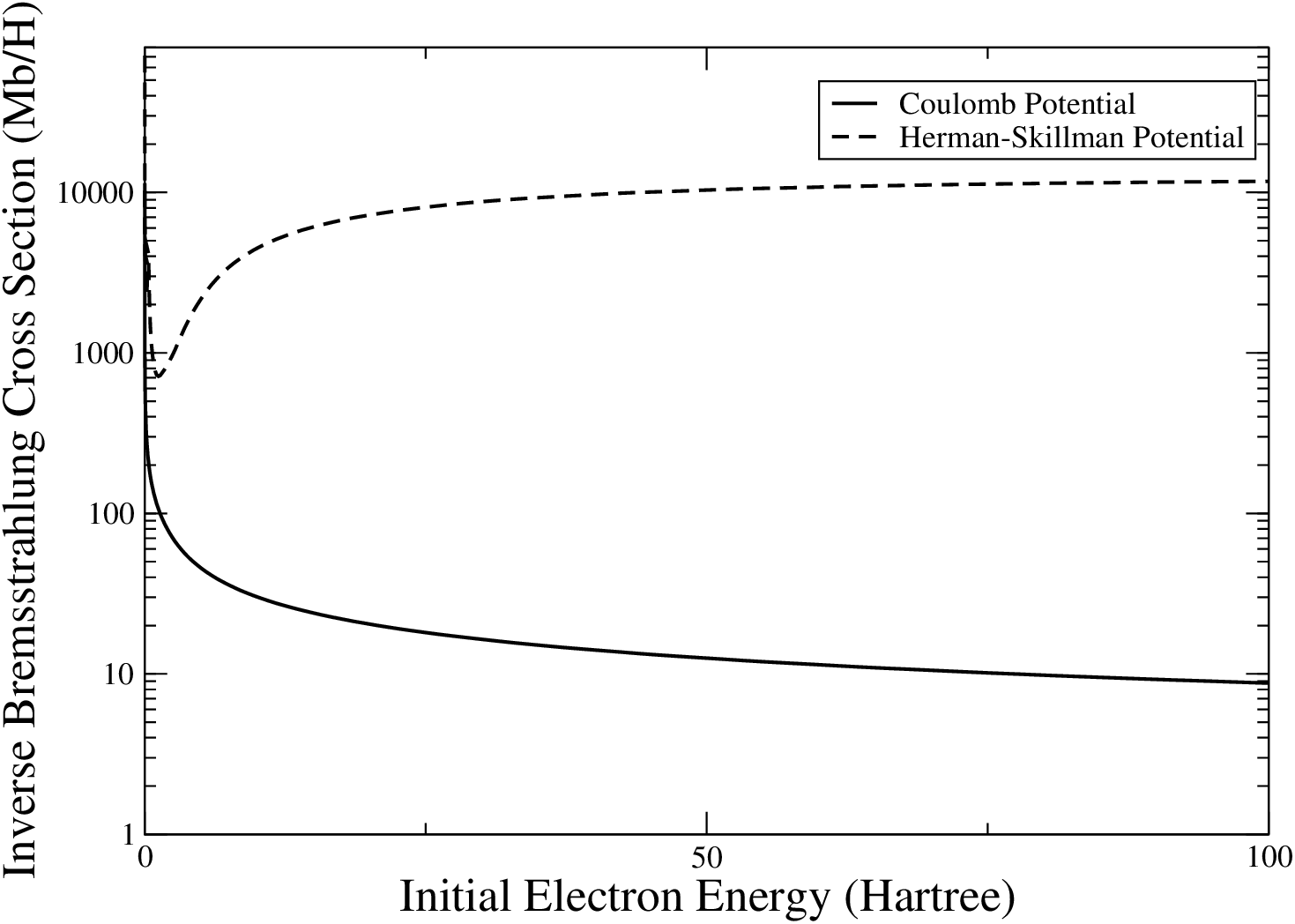}
	\caption{Inverse bremsstrahlung cross sections [Eqs. \ref{IBS1} and \ref{IBS2}] 
for an electron with incident energy $E$ to absorb a $12.7$-eV photon are given for an  
electron in the field of a purely Coulombic $1+$ potential and for an electron in the field  
of a Xe Herman-Skillman atomic potential. The effects of atomic structure on inverse 
bremsstrahlung rates are quite pronounced.} 
	\label{fig:IBSComparison} 
	\end{center} 
\end{figure}

\begin{figure} 
	\begin{center} 
	\includegraphics[width=3.375in]{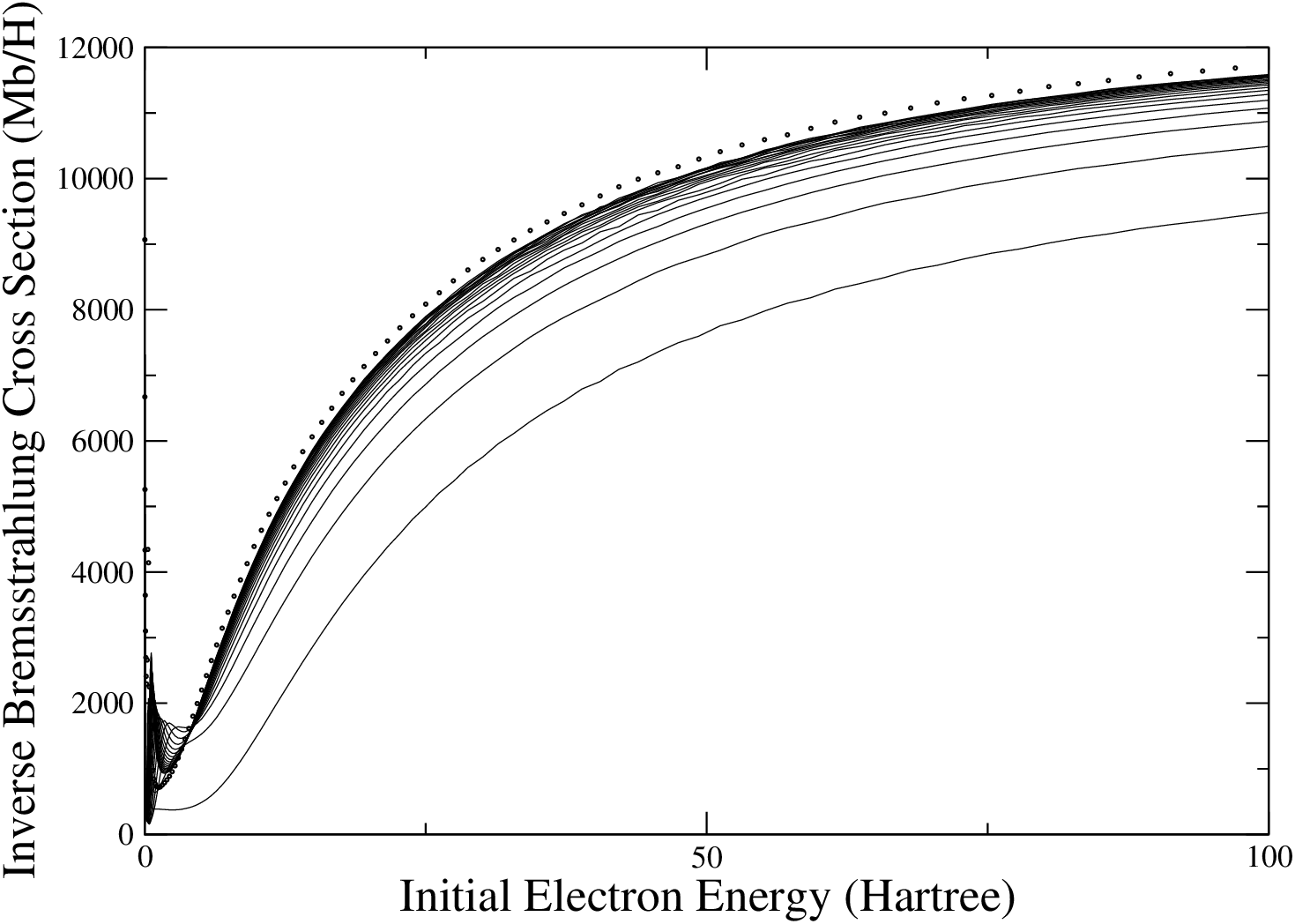}
	\caption{The inverse bremsstrahlung cross section as a function of  
energy is shown for an electron in the field of a Debye-screened Xe 
Herman-Skillman potential, with the Debye screening length $\lambda_D$ ranging from 1 a.u. to 20 a.u.   
As $\lambda_D$ grows, the cross section approaches the limit of no plasma  
screening, shown in this graph by the dotted line. As the Debye length of the cluster  
plasma shrinks, the charged ion is shielded more effectively from the scattering electron, 
and the inverse bremsstrahlung cross section is decreased.} 
	\label{fig:IBSvsE} 
	\end{center} 
\end{figure}

Figures \ref{fig:IBSComparison} and \ref{fig:IBSvsE} illustrate the 
dramatic effects of the ionic potential on the inverse bremsstrahlung 
cross section. In Fig. \ref{fig:IBSComparison}, as the scattering 
electron collides with the ion at 
higher and higher initial energies, it 
probes regions of the ionic potential at which the ion nucleus is 
screened increasingly poorly by inner-shell electrons.  As a 
result, the inverse bremsstrahlung cross section rises to many 
hundreds of times that of the naked Coulomb potential. 
 
Adding plasma screening to this picture has the effect of 
supplementing the screening effects of inner-shell electrons with the 
screening effects of plasma electrons.  As a result, the scattering 
electron feels the effects of the ionic nucleus more 
strongly than in the pure Coulomb case, but less strongly 
than in the case of the unscreened ionic potential.  This is 
seen in a steady decrease of the inverse bremsstrahlung cross section 
as the screening range decreases.

\begin{figure}%figure updated 
	\begin{center} 
	\subfigure{ 
	 \includegraphics[width=3.375in]{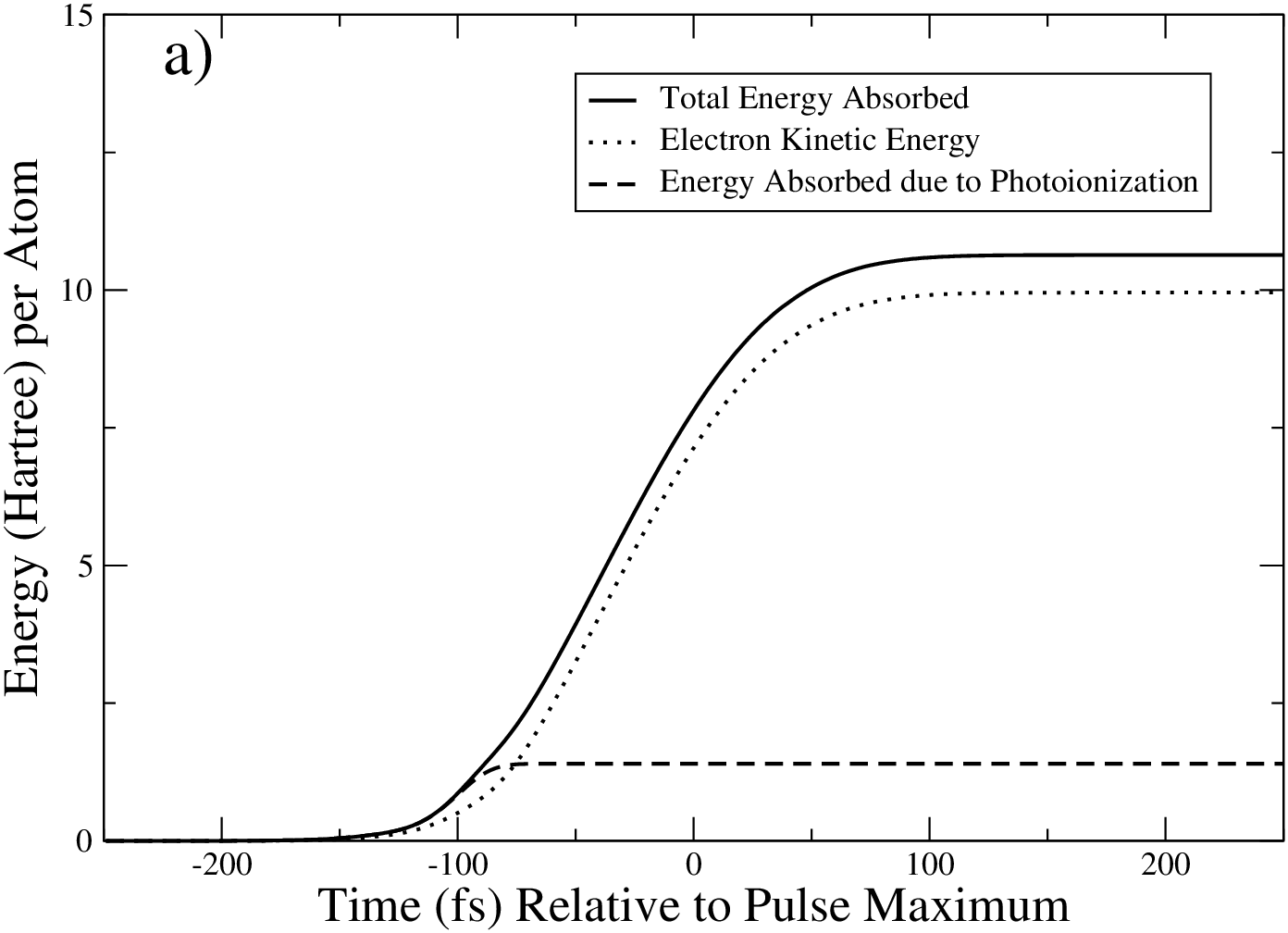}}
\vskip 0.15in 
	 
	\subfigure{ 
	\includegraphics[width=3.375in]{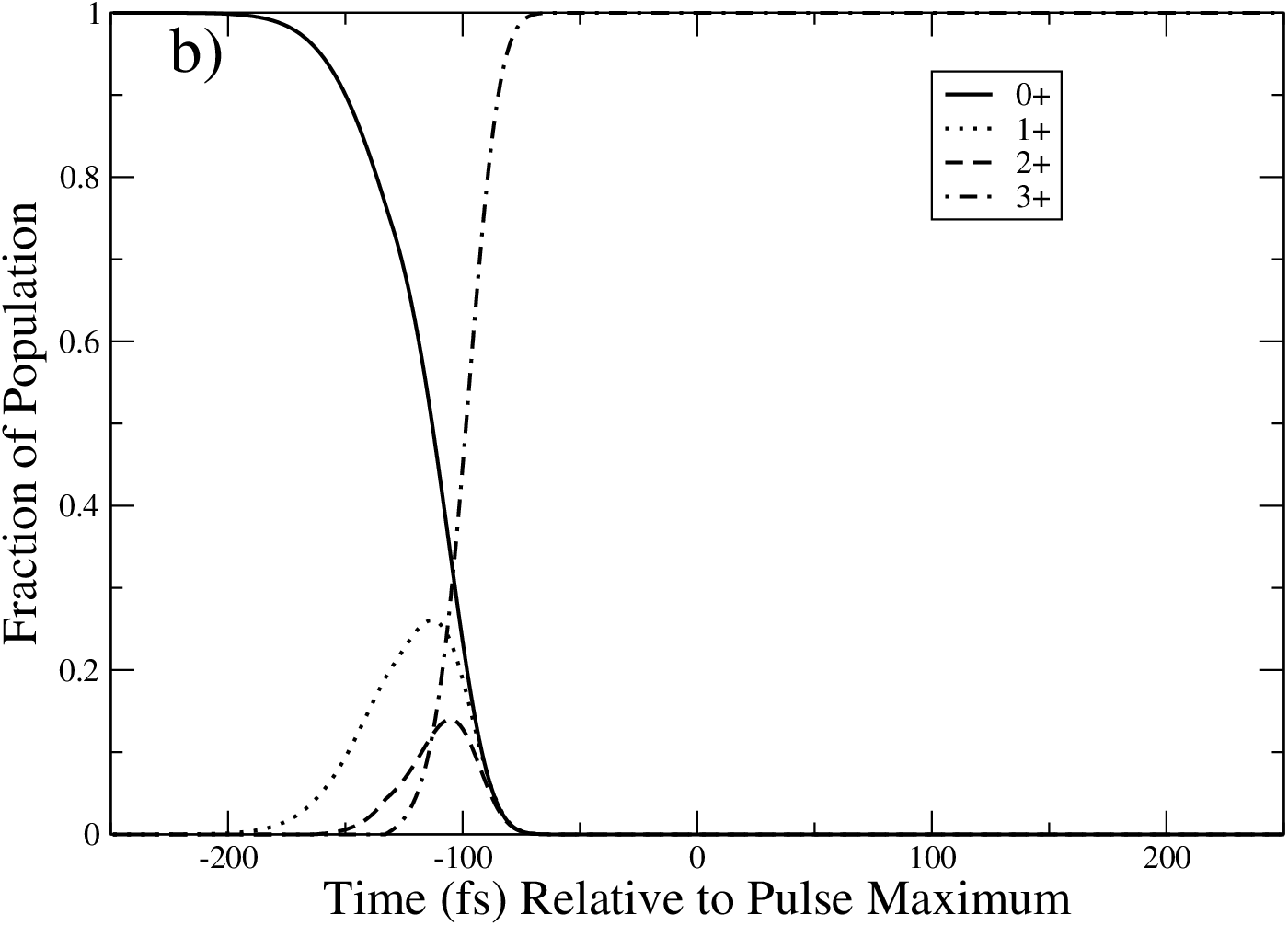}} 
	\end{center} 
	\caption{Evolution of a 1500 atom cluster exposed to  a 100
fs, $7 \times 10^{13}$ W/cm$^{2}$ pulse, employing only 
photoionization and inverse bremsstrahlung heating.  
a) Energy absorbed vs. time.
b) Ionic population vs. time. Xe$^{2+}$ and 
Xe$^{3+}$ are produced efficiently via photoionization.} 
	\label{fig:PIonly} 
\end{figure} 
 
\section{Collisional Ionization and Recombination} 
 
Although photoionization and inverse bremsstrahlung are the only 
processes by which the cluster can absorb photons from the laser 
beam, they are not by themselves sufficient to explain the cluster's 
evolution.  As the pulse progresses, large numbers of free 
electrons fill the cluster.  These electrons can liberate other 
electrons via collisional ionization if they have sufficient energy, 
or they can undergo three-body recombination with an ion. 
 
Including the effects of ionization and recombination, the rate 
equation for the number per unit volume $n_{i}$ of charge species $i$ 
is given by  
\begin{eqnarray} 
\frac{\partial n_{i}}{\partial t} & = &\frac{I}{\omega}(\sigma_{\text{PI}}^{i-1} \, n_{i-1}- 
\sigma_{PI}^{i} \, n_{i})  \nonumber \\ 
& & + S_{i-1}n_{i-1}n_{e}-S_{i}n_{i}n_{e} \\ 
& & + R_{i+1}n_{i+1}n_{e}^2-R_{i}n_{i}n_{e}^2 \; , \nonumber 
\end{eqnarray} 
where $n_{e}$ is the number of electrons per unit volume.   The photoionization 
cross sections $\sigma_{\text{PI}}^{i}$ were calculated  
in Section \ref{sec:Photoionization}. 
The ionization and recombination coefficients $S_{i}$ and $R_{i}$ for the 
reaction Xe$^{i+}$+e$^{-} \rightarrow$Xe$^{(i+1)+}$+2e$^{-}$  are 
calculated later in this section.

These rate equations, along with equations for the energy in the free 
electron gas and the radius of the cluster are integrated numerically 
through the duration of the pulse.  As a general rule, 
this set of equations will be quite stiff.  We performed this 
integration using the Rosenbrock method \cite{NumRec}.

There are two requirements for a satisfactory treatment of collisional 
ionization and recombination in the cluster.  First, both processes 
must occur at appropriate rates, and second, the rates for ionization 
and recombination must be consistent with one another, in that they 
drive the cluster towards chemical equilibrium at all times. 
The second requirement is particularly significant because the usual 
treatment of collisional ionization (also used in this study) uses the 
semiempirical Lotz formula \cite{lotz} for ionization from the 
$j^{\text{th}}$ subshell, 
\begin{eqnarray} 
S^{j}_{i} & = & 6.7 \times 10^{-7}\frac{a_j q^{j}_i}{T^{3/2}} 
\left(\frac{1}{P_{j}/T}\int_{P_{j}/T}^{\infty} \frac{e^{-x}}{x} dx \right. \\ 
& & \left. -\frac{b_{j} \exp{(c_{j})}}{P_{j}/T+c_{j}}\int_{P_{j}/T+c_{j}}^{\infty} 
\frac{e^{-y}}{y} dy\right) \frac{\mathrm{cm}^3}{\mathrm{s}} \; , \nonumber 
\label{lotzformula} 
\end{eqnarray} 
to find ionization coefficients, where $a_{j}, b_{j}, c_{j}$ are 
semiempirical constants, $q^{j}_{i}$ the number of equivalent electrons 
Xe$^{i+}$ contains in the $j^{\text{th}}$ subshell, $P_{j}$ the 
ionization potential, and $T$ the temperature.  For charge states of 
$0,\ldots,5+$, we choose semiempirical constants corresponding to the 5p 
sublevel.  For charge states of $6+$ and $7+$, which have no 5p electrons 
in the ground state, we choose constants corresponding to the 5s sublevel. 
Because this formula is only a semiempirical approximation, 
it is important to use recombination coefficients 
which are consistent with the ionization coefficients to prevent the 
model from settling into an incorrect equilibrium charge distribution. 
 
The ratio between ionization and recombination coefficients 
can be obtained using the concept of equilibrium constants.  In a 
plasma at equilibrium, the rate of collisions ionizing Xe$^{i+}$ to 
form Xe$^{(i+1)+} + e^{-}$ must be equal to the rate at which 
Xe$^{(i+1)+} + e^{-}$ recombines to form Xe$^{i+}$.  The recombination 
coefficients found in this manner can then be applied to modeling the 
cluster plasma, which is not, in general, in a state of chemical equilibrium. 
 
These two rates are given respectively by 
\begin{equation} 
S_{i} n_{i} n_{e} = \text{rate of ionizing collisions} 
\label{IonizationRate} 
\end{equation} 
and 
\begin{equation} 
R_{i+1}n_{i+1}n_{e}^2 = \text{rate of recombining collisions} \; . 
\label{RecombinationRate} 
\end{equation} 
Hence, the ratio $S/R$ is given by 
\begin{equation} 
\frac{S_{i}}{R_{i+1}} = \frac{n_{i+1}^{eq}n_{e}^{eq}}{n_{i}^{eq}} \; . 
\label{RecombRatio} 
\end{equation} 
The fraction $(n_{i+1}^{eq}n_{e}^{eq})/n_{i}^{eq}$ is known as the 
equilibrium constant for the reaction, and can be calculated thermodynamically. 
 
In any reaction $A \rightarrow B + C$ at equilibrium, the chemical 
potentials for the forwards and backwards reactions must be balanced 
$\mu_A = \mu_B+\mu_C$.  The chemical potential of each species is 
given by a partial derivative of the Helmholtz free energy,  
\begin{equation} 
\mu_{i}=\frac{\partial F}{\partial N_{i}}_{T,V} \; . 
\end{equation} 
The Helmholtz free energy is given by $F=-T \ln Z_{tot}$, where $Z_{tot}$ is the partition 
function for the system as a whole. 
 
Factoring the total partition function into the product of individual 
particle partition functions (which implies independent particles),  
\begin{widetext} 
\begin{equation} 
Z_{tot}(N_A,N_B,N_C,V,T)= 
Z_A(N_A,V,T)Z_B(N_B,V,T)Z_C(N_C,V,T)= 
\frac{z_A(V,T)^{N_A}}{N_A!}\frac{z_B(V,T)^{N_B}}{N_B!} 
\frac{z_C(V,T)^{N_C}}{N_C!} \; , 
\label{TotalZ} 
\end{equation} 
\end{widetext} 
and assuming $N_{i} >> 1$ yields 
\begin{equation} 
\mu_i=-T \frac{\partial \ln(Z_i(V,T))}{\partial N_i}= 
-T \ln(\frac{z_i}{N_i}) \; . 
\label{ChemPot2} 
\end{equation} 
Imposing balanced chemical potentials yields  
\begin{equation} 
\frac{N_B N_C}{N_A}=\frac{z_B(V,T) z_C(V,T)}{z_A(V,T)} 
\label{EqConst1} 
\end{equation} 
or equivalently 
\begin{equation} 
K_{\text{eq}}(T)= 
\frac{n_{B}n_{C}}{n_{A}}=\frac{\frac{z_B(V,T)}{V}\frac{z_C(V,T)}{V}} 
{\frac{z_A(V,T)}{V}} \; , 
\label{EqConst2} 
\end{equation} 
where  the equilibrium constant $K_{\text{eq}}$ is a function of temperature 
only. 
 
If the ionization potential of Xe$^{i+}$ is given  
by $P_{i}$, then the partition functions are given by 
\begin{eqnarray} 
z_i & = & \int_{0}^{\infty} d E e^{-E/T}\rho_{i}(E) \; , \nonumber \\ 
z_{i+1} & = & e^{-P_{i}/T} \int_{0}^{\infty}d E e^{-E/T}\rho_{i+1}(E) \; , \\ 
z_{e} & = & \int_{0}^{\infty}d E e^{-E/T}\rho_{e}(E) \; . \nonumber 
\end{eqnarray} 
Through most of the lifetime of the pulse, tight plasma screening 
destroys the Rydberg states  and most of the internal degrees of 
freedom of the various ions, leaving the density of states $\rho(E)$ 
dominated by the center of mass term and by a combinatorial term
\begin{equation}
D(i) =\begin{pmatrix}m \\ n \end{pmatrix}
\end{equation} 
accounting for  the number of ways $n$ electrons 
can be distributed in $m$ orbitals. For charge states up to $6+$, we
use $m=6, n=6-i$.  For $7+$ and $8+$, we use $m=2$,$n=8-i$.
If we exploit this by setting 
$\rho_i(E)/D(i)=\rho_{i+1}(E)/D(i+1)$, the common integral in $z_i$
and $z_{i+1}$  
falls out of the equilibrium constant, yielding 
\begin{equation} 
K_{\text{eq}}^{(Xe^{i+} \rightarrow Xe^{(i+1)+}+e^{-})}= 
e^{-P_{i}/T} 
\frac{T^{\frac{3}{2}} D(i+1)} 
{\sqrt{2} \pi^{\frac{3}{2}} D(i)} 
\label{EqConst3} 
\end{equation} 
Equation (\ref{EqConst3}) can now be combined with Eqs. 
(\ref{lotzformula}) and (\ref{RecombRatio}) to yield recombination rate 
coefficients which have appropriate magnitude and which, in 
combination with the ionization coefficients, drive the system toward 
the correct equilibrium distribution at all times. 
 
A gas of charged particles has different thermodynamic properties from 
an ideal gas due to Coulomb interactions between the constituent 
particles.  Zel'dovich and Raizer \cite{zeldovich} calculate the 
adjustment to $K_{\text{eq}}$ due to a Debye-H\"uckel potential. 
The equilibrium constant including Coulomb effects can be 
written  
\begin{equation} 
K_{\text{eq}}^{(Xe^{i+} \rightarrow Xe^{(i+1)+}+e^{-})}= 
e^{-(P_{i}+\Delta P_{i})/T} 
\frac{T^{\frac{3}{2}} D(i+1)} 
{\sqrt{2} \pi^{\frac{3}{2}} D(i)} \; , 
\end{equation} 
where the change in ionization potential due to Coulomb effects is  
$\Delta P_{i}=-(Q_{i}+1)/\lambda_{D}$, the Coulomb potential between  
the ion core and an electron held at distance $\lambda_{D}$. 
 
In our approach, by explicitly calculating bound state energies for 
Debye-screened Hartree-Slater potentials, we calculate this adjustment to the 
ionization potential directly.  Our adjustment behaves similarly to 
the Zel'dovich and Raizer correction, but is larger for longer 
screening lengths and smaller at shorter screening lengths. 
 
One advantage to the equilibrium constant approach is that it conceptually 
separates information about thermodynamic 
balance from the rate at which the system seeks that balance.  As a 
result, any formula for ionization or recombination coefficients could 
be substituted for the Lotz formula, with the accuracies of the overall 
rate and of the equilibrium constant used 
the only criteria for validity of the formula. 
%This advantage is also seen in the formula by Hahn \cite{Hahn97} 
%which we employ in Sec. \ref{SCHE} in calculating collisional 
%ionization and recombination during the cluster's expansion. 
 
\begin{figure}%%figure updated 
	\begin{center} 
	\subfigure{ 
	\includegraphics[width=3.375in]{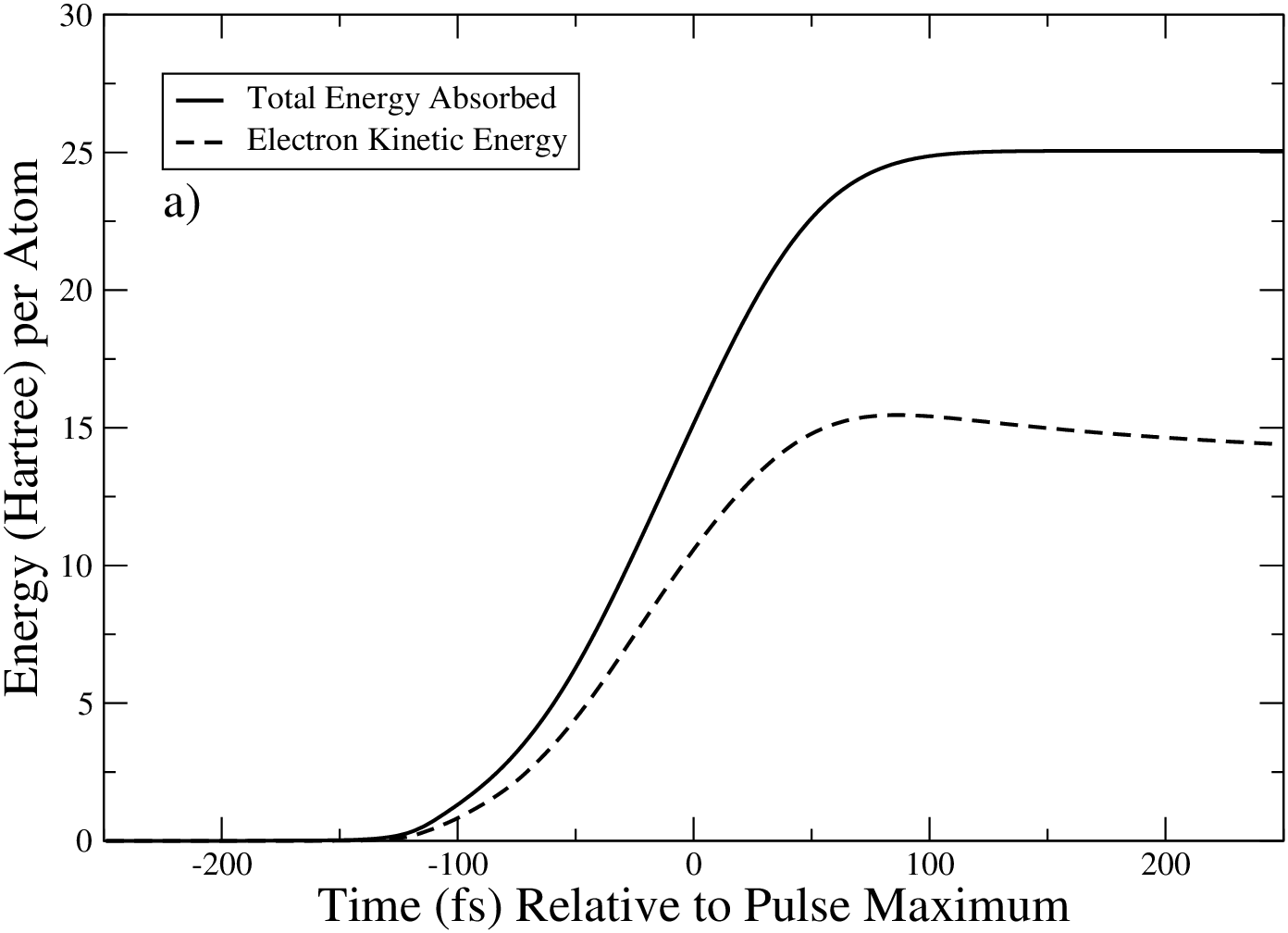}} 
	\vskip 0.15in 
	\subfigure{ 
	\includegraphics[width=3.375in]{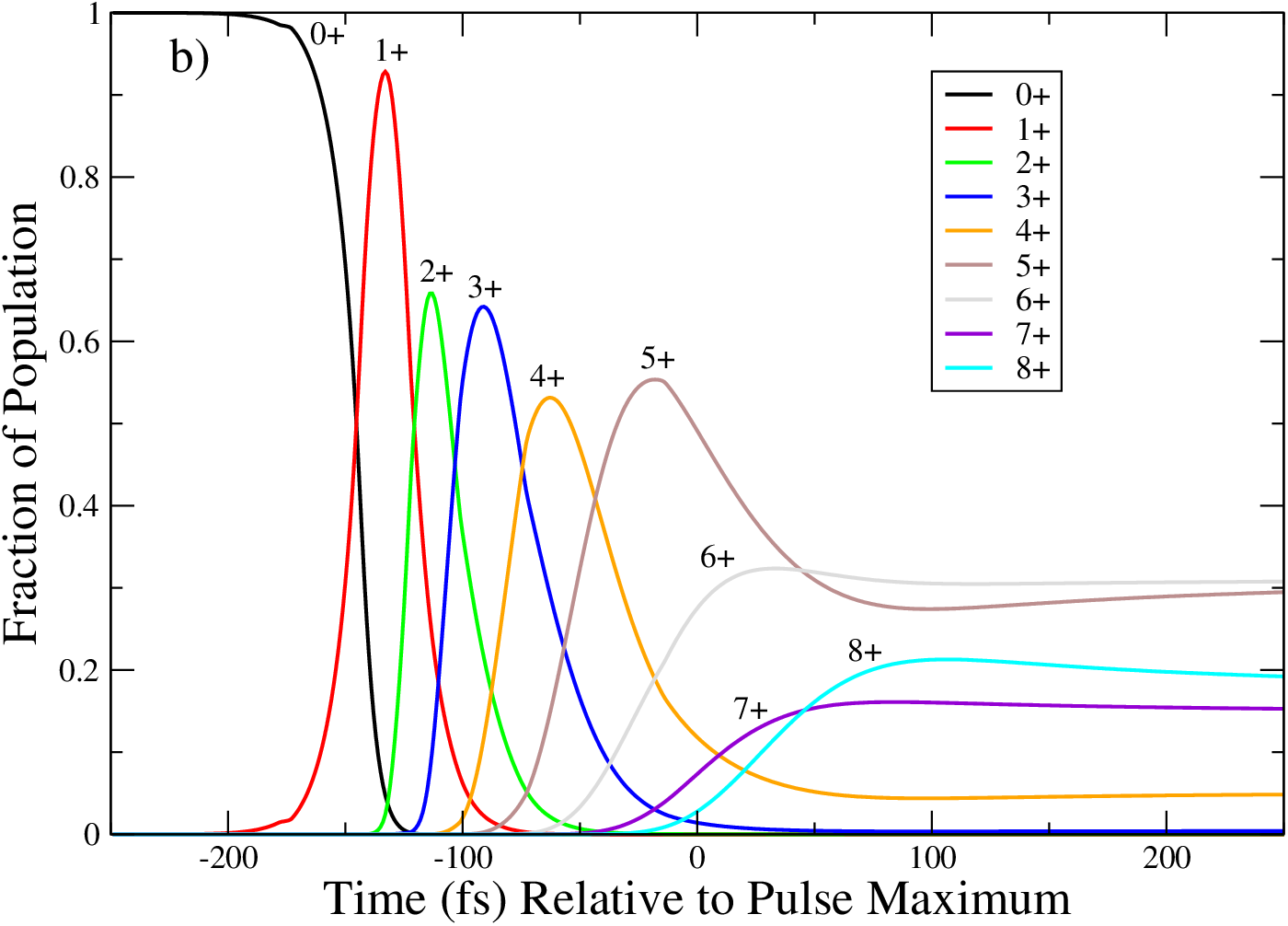}} 
	%\vskip 0.15in 
	%\subfigure{ 
		%\includegraphics[width=3.375in]{natureparameters_debyeWigSeitz_chargepop_1d5}}
	\end{center} 
	\caption{The effects of collisional ionization and 
recombination are to allow the formation of charge states beyond 
Xe$^{3+}$  Pictured is the time evolution of a single 1500 atom
cluster exposed to a 100 fs, $7 \times 10^{13}$ W/cm$^2$ pulse.
These states enhance the 
rate of inverse bremsstrahlung heating.   
As the plasma expands and cools, the 
chemical equilibrium shifts toward lower charge states on a timescale 
much longer than the laser pulse, until decreasing plasma density 
causes recombination and ionization rates to go to zero. a) Energy 
absorbed vs. time b) Ionic population vs. time during laser pulse}
%c) Ionic population vs. time during cluster expansion  } 
	\label{fig:natureparams_debyeWigSeitz} 
\end{figure}

Including the effects of collisional ionization and recombination has 
a pronounced effect on the evolution of the cluster.  In Fig. 
\ref{fig:PIonly}, the evolution of the cluster is calculated employing 
only photoionization and inverse bremsstrahlung.  In contrast, Fig. 
\ref{fig:natureparams_debyeWigSeitz} shows the evolution of the same 
cluster employing photoionization, inverse bremsstrahlung, collisional 
ionization and recombination, and evaporation of energetic electrons 
from the cluster.  Allowing ionization and recombination has the 
effect of producing charge states up to Xe$^{8+}$ in substantial quantities, 
and of nearly doubling the energy per atom absorbed by the cluster.

\section{Cluster Dynamics During the Laser Pulse} 
 
As the cluster absorbs energy from the laser field, some of the 
electrons become so energetic that they are no longer bound to the 
cluster.  In addition, the cluster expands and cools due to 
hydrostatic forces from the hot electrons and Coulomb repulsion as 
escaping electrons leave a charge imbalance behind.  These in turn 
affect the microscopic processes inside the cluster, since all such 
processes depend on the concentrations of charge species within the 
cluster.  Collisional ionization and recombination are also 
sensitively dependent on the temperature of the electron gas relative 
to electron binding energy. 
 
For the evolution of the cluster during the period of the laser pulse, 
we employed a simple model \cite{ditmire96} of the cluster expansion 
which tracks only the radius of the cluster, the evaporation of 
electrons away from the cluster, and the loss of heat from the 
electron gas accompanying 
both processes.  We did not consider the possibility of either 
gross movement of electrons or spatial inhomogeneity of charge species 
within the cluster, processes which a recent theoretical 
study \cite{siedschlag} has suggested may account for the formation of 
highly charged ions detected at the Hamburg experiment.   
 
The equation for the radius of the cluster is given by 
\begin{equation} 
\frac{\partial^2 r}{\partial t^2}=3\frac{P_e+P_{\text{Coul}}}{n_{\text{Xe}} 
m_{\text{Xe}}} \frac{1}{r} \; , 
\end{equation} 
where $P_e=n_e T_e$ is the electron pressure and  
$P_{\text{Coul}}=Q^2/(8 \pi r^4)$ is the Coulomb pressure 
resulting from the charge built up as electrons evaporate away from 
the cluster. 
 
This model of the laser--cluster dynamics also distinguishes between 
inner and outer ionization.  Inner ionization, which takes place due 
to photoionization and collisional ionization, is the process by which 
electrons become liberated from their parent ion and join the cluster 
plasma, where they can undergo inverse bremsstrahlung heating or 
collisional ionization/recombination.  Outer ionization is the process 
by which electrons with sufficient energy escape the cluster and cease 
to have interactions with it. 
 
The rate of evaporation from a Maxwell distribution of electrons can 
be calculated knowing the size of the cluster, the mean free path of 
electrons in the cluster, and the temperature of the electron plasma. 
The rate at which electrons escape from the cluster is then given by 
\begin{equation} 
W_{fs}=\int_{v_{esc}}^{\infty} d v \frac{\pi}{4} \frac{\lambda_{e}}{r}(12 r^2-\lambda_{e}^{2}) 
v f(v) 
\label{WFS} 
\end{equation} 
where $v_{esc}=\sqrt{2(Q+1)/r}$ is the velocity required for an 
electron to escape from a cluster of charge $Q$,  
\[f(v)= 4 \pi n_{e} (2 \pi T)^{-3/2} v^{2} e^{-\frac{v^{2}}{2 T}}\] is 
the Maxwell distribution, and $\lambda_{e}$ is the mean free path in 
the cluster plasma, given by  
\[\lambda_{e}=\frac{T^{2}}{4 \pi n_{e} (Z+1) \ln{\Lambda}}\]  
for a plasma with average ion charge $Z$. The Coulomb logarithm, 
$\ln{\Lambda}$, is set equal to $1$ in our calculation of the mean free path.  
$\lambda_{e}$ is constrained to be no greater than $2 r$, the diameter
of the sphere. 
 
As electrons evaporate from the cluster, the remaining cluster becomes 
ever more highly charged, and a correspondingly lower fraction of the 
Maxwell distribution has enough energy to escape the cluster, thereby 
choking off the evaporation rate.   
 
It is likely that nearly all high-energy electrons detected in the
experiment escape  
during this original period of evaporation.  As the cluster expands,  
the temperature of the electron plasma falls very quickly as electron thermal  
energy is converted into ion kinetic energy, while the energy required
to escape the cluster falls only as $1/r$. 
 
A recent experiment \cite{laarman05} has for the first time measured 
the energy spectrum for electrons emitted from rare gas clusters 
exposed to intense VUV light.  They give ejection spectra for  
70 atom xenon clusters 
exposed to a $4.4 \times 10^{12}$~W/cm$^2$ pulse of VUV light at the 
same photon energy as the original Hamburg experiment, finding an electron 
distribution which decreases approximately exponentially according to 
$I=I_{0}\exp{(-E_{\text{kin}}/E_{0})}$, with $E_{0}= 8.9$ eV.   
 
\begin{figure}%%figure updated 
	\begin{center} 
	\includegraphics[width=3.375in]{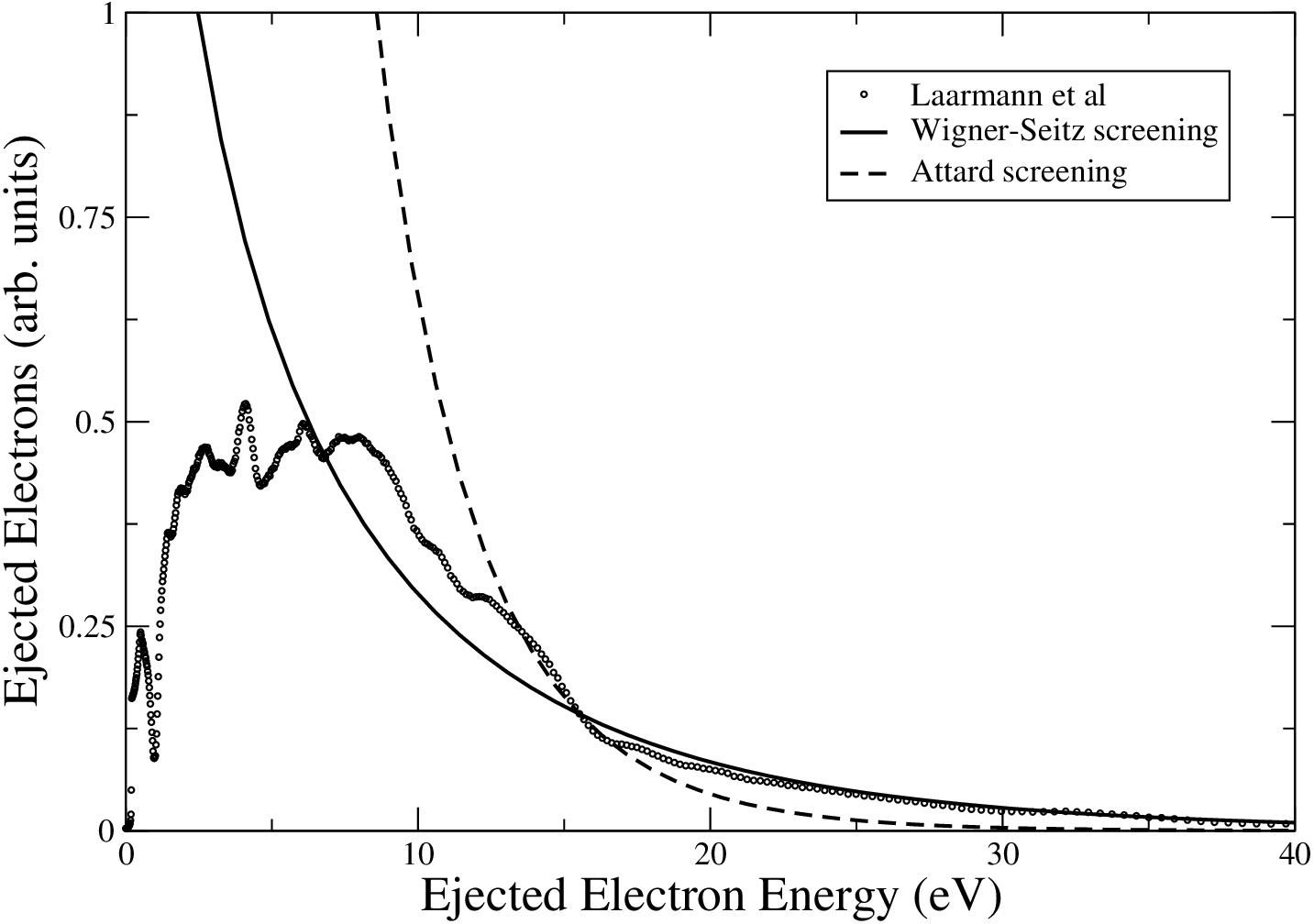}
	\end{center} 
	\caption{Ejected electron spectrum. Comparison between data
from \cite{laarman05} and spatially-averaged 
spectra calculated using 70 atom clusters exposed to a $4.4 \times 10^{12}$ 
W/cm$^2$, 100 fs pulse for two different models of plasma screening.
The Wigner Seitz cutoff model uses the ordinary Debye length as the
screening radius, but the screening radius is not allowed to fall
below xenon's Wigner-Seitz radius at liquid density, 4.64 bohr.  The
Attard model of screening calculates the screening radius according to
equation (\ref{attardscreening}), discussed in Section
\ref{sec:nonidealscreening}. 
The spectrum calculated using xenon's Wigner-Seitz radius as a minimum 
screening distance displays a strong similarity to the experimental
curve.  The intensity of the experimental spectra is arbitrary; magnitudes
were chosen by setting each curve equal at the beginning of the
exponential tail in the experiment.}
	\label{fig:PRLejectedelectrons} 
\end{figure}

We calculated a spectrum of ejected electrons by stepping through a 
laser pulse using small timesteps.  For each timestep, we calculated 
the electron density, mean free path, cluster radius, and plasma 
temperature.  Using these parameters, we calculated the rate at which 
electrons with energy $E=E_{\text{esc}}+E_{\text{kin}}$ escaped from 
the cluster using Eq. (\ref{WFS}).  Integrated through the timescale of a 
pulse until the evaporation has stopped, this yields an ejected 
electron spectrum for a single cluster exposed to the pulse.  Since 
the clusters are located randomly with respect to the center of the 
laser pulse, we further performed a spatial integration over the 
radial dimension of the pulse, assuming a Gaussian  laser profile
$I(r) \propto e^{-r^2/\sigma^2}$ from 0 to 
3 $\sigma$.  The length of the interaction region in the 
Hamburg experiment was comparable to the Rayleigh range for the laser; 
accordingly, we assumed a constant laser intensity along the direction 
of propagation. After performing the spatial integration, we found that
on average .22 electrons per xenon atom evaporated from the cluster in
this way.
The spectrum of ejection energies for these electrons shown in Fig. 
\ref{fig:PRLejectedelectrons}, although not exponential, is 
nevertheless quite similar to the electron spectrum found in
Ref. \cite{laarman05}. 
 
The largest discrepancy between our calculated spectrum and the 
spectrum from \cite{laarman05} occurs at low ejection energy.  In 
addition, our model of the cluster expansion predicts that the 
majority of electrons will comprise electron plasmas which remain 
bound to the cluster ions and become quite cold during the process of 
expansion.  These electrons would reach the detector at low energies 
and after long delay times, further boosting the spectrum at low energies. 
However, Laarmann {\em et al.} note that for $E_{\text{kin}}<2.5$ eV, 
coinciding with the region of largest discrepancy, the spectrum cannot 
be evaluated due to large levels of noise in the background spectra. 
 
\begin{figure}%%figure updated 
	\begin{center} 
	\subfigure{ 
	\includegraphics[width=3.375in]{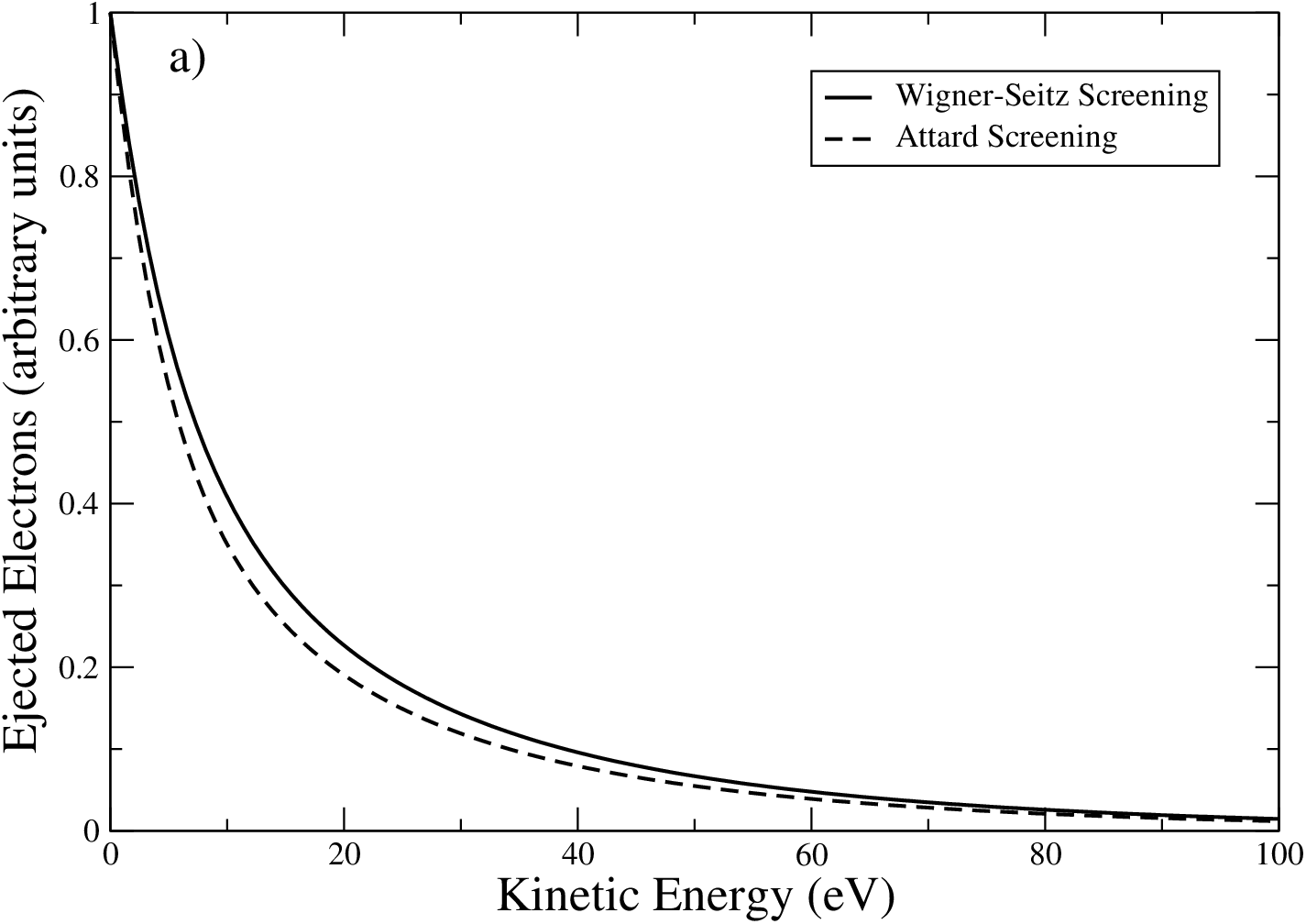}} 
	\vskip 0.15in 
	\subfigure{ 
	\includegraphics[width=3.375in]{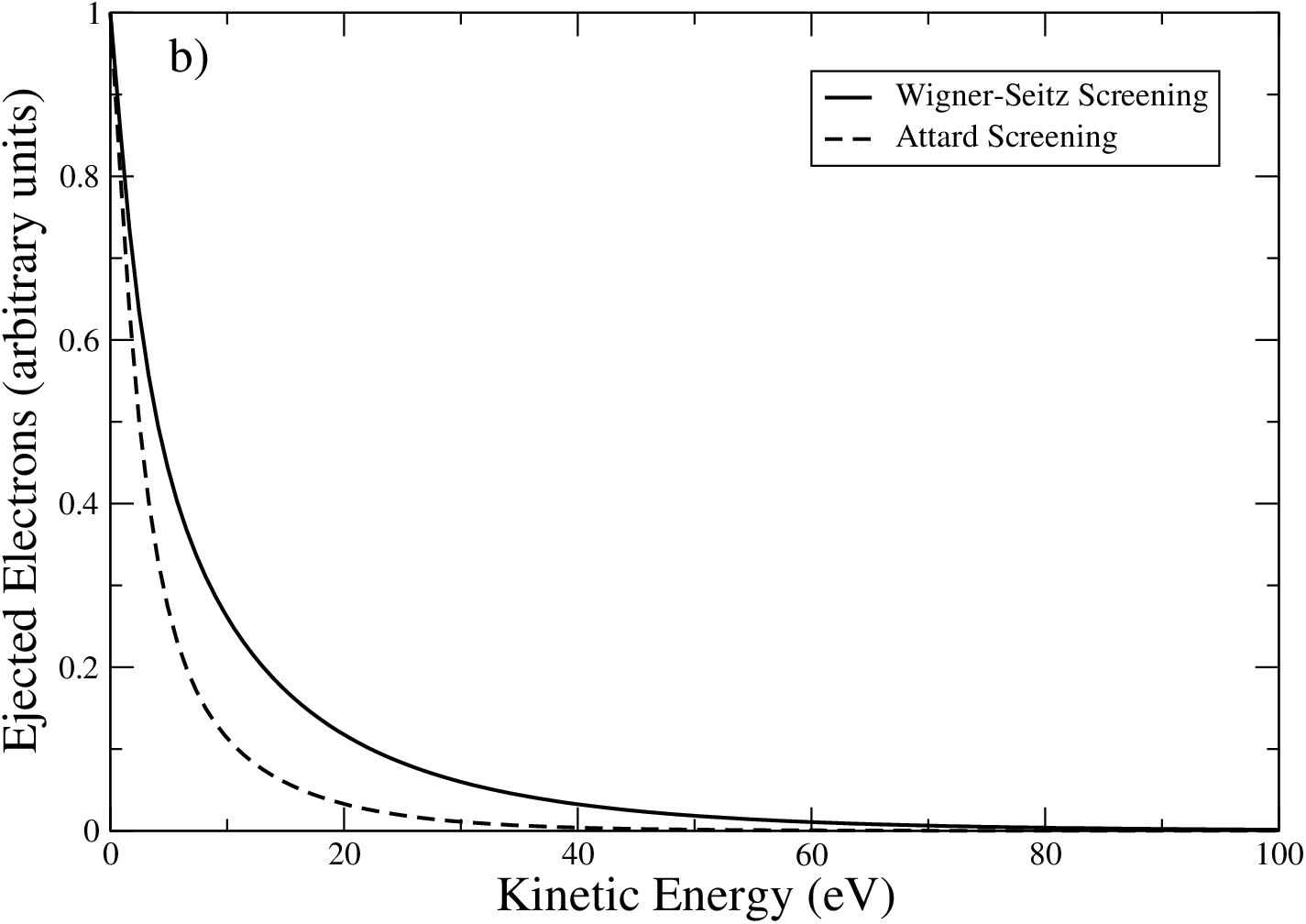}}
	\end{center} 
	\caption{Ejected electron spectra, calculated for the two sets
of parameters and the two models of screening. 
The Wigner Seitz cutoff model uses the ordinary Debye length as the
screening radius, but the screening radius is not allowed to fall
below xenon's Wigner-Seitz radius at liquid density, 4.64 bohr.  The
Attard model of screening calculates the screening radius according to
equation (\ref{attardscreening}), discussed in section
\ref{sec:nonidealscreening}. 
a) Nature parameters: 
1500 atom clusters exposed to a 100 fs, $7 \times 10^{13}$ W/cm$^{2}$ 
pulse. b) Thesis parameters: 2500 atom clusters exposed to a 50 fs,  
$2.5 \times 10^{13}$ W/cm$^{2}$ pulse. 
Since electrons
faster than about 1 eV are ejected from the cluster during the pulse
the  ejection spectra could serve as a window into the
dynamics of the  laser-cluster interaction.} 
	\label{fig:ejectedelectrons} 
\end{figure}

Since electrons
faster than about 1 eV are ejected from the cluster during the pulse
rather than during the slower process of 
cluster expansion, the ejected electron spectrum has the potential to 
serve as a window into the nature of the laser--cluster interaction. 
Accordingly, we give the spectra for 1500 atom clusters exposed to a 100 
fs, $7 \times 10^{13}$ W/cm$^{2}$ pulse, and for 2500 atom clusters 
exposed to a 50 fs, $2.5 \times 10^{13}$ W/cm$^{2}$ pulse in 
Fig. \ref{fig:ejectedelectrons}.

After spatial averaging, we find that 1500 atom clusters exposed to a
100 fs, $7 \times 10^{13}$ W/cm$^{2}$ pulse eject 0.22 electrons per
atom during this early evaporation period using the Wigner-Seitz
cutoff model for the screening length (see section
\ref{sec:nonidealscreening} for a discussion of plasma screening).
Using the Attard model, 0.07 electrons per atom are evaporated during
this period.  For  2500 atom clusters exposed to a 50 fs, $2.5 \times
10^{13}$ W/cm$^{2}$ pulse, the corresponding numbers are 0.13 electrons
per atom for the Wigner-Seitz cutoff model and 0.02 electrons per atom
for the Attard model.  In contrast to this, the Hamburg experiment
measured an average charge per ion of 2.98.  Hence, the electrons
which comprise these ejected electron spectra correspond to only a few
percent of all free electrons at the time when the expanding clusters
reach the detector.

%\section{Spatial Averaging} 
% 
%Each laser pulse interacts with several clusters which are distributed 
%randomly with respect to the focal point of the pulse.   
%Assuming a gaussian laser profile $I(r) \propto e^{-r^2/\sigma^2}$, 
%we integrate the charge state population and ion energy over the range 
%$r=0$ to $3 \sigma$.  The length of the interaction region in the 
%Hamburg experiment was comparable to the Rayleigh range for the laser; 
%accordingly, we assumed a constant laser intensity along the direction 
%of propagation. 
 
\section{Nonideal Plasma Screening}\label{sec:nonidealscreening} 
 
\begin{figure}%%figure updated 
	\begin{center} 
	\includegraphics[width=3.375in]{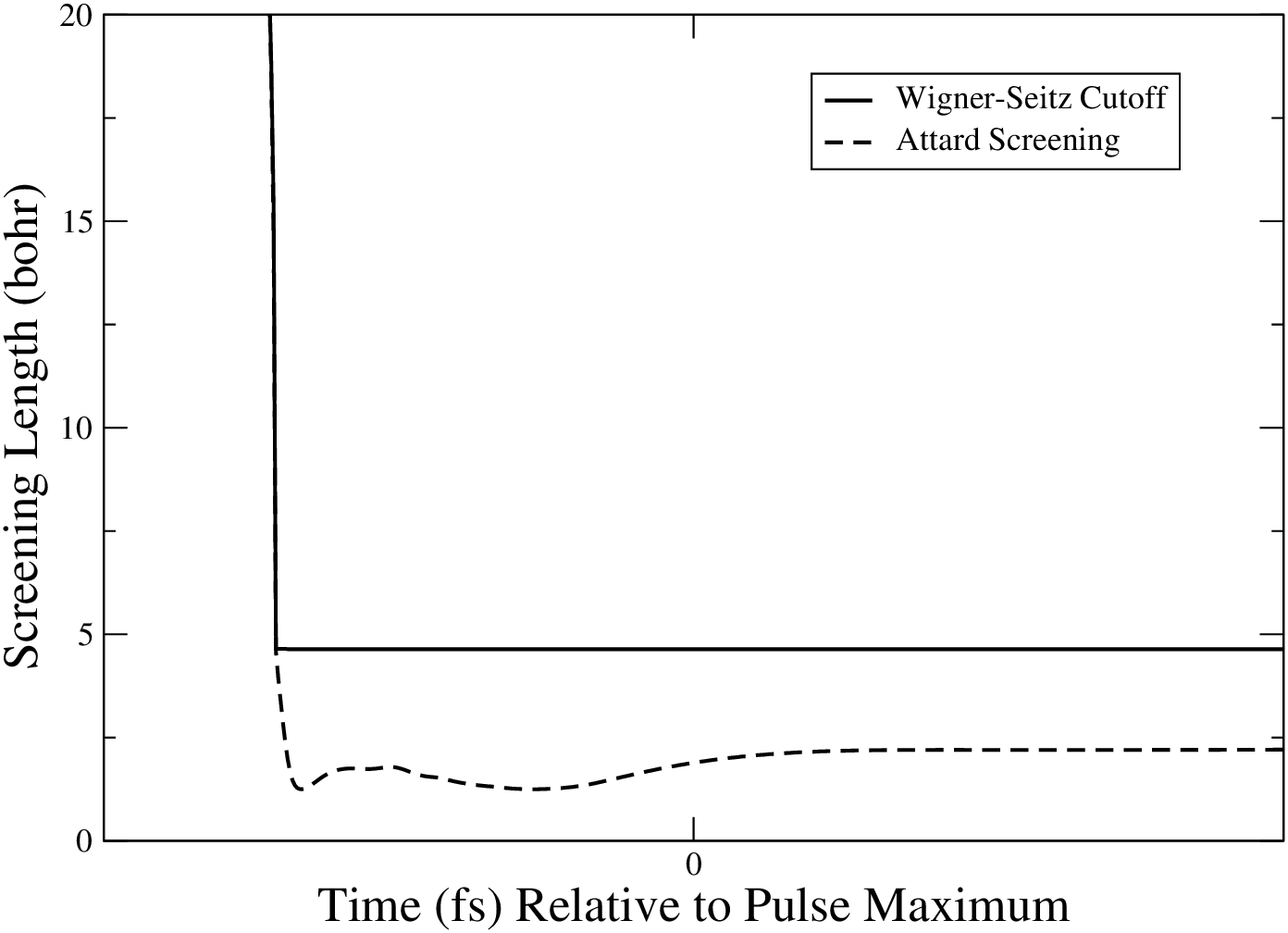} 
	\end{center} 
	\caption{The interaction of plasma screening with atomic 
potentials is unknown as the screening length becomes very short. 
Here the screening length vs time is given for two simulations of
a 1500 atom cluster 
exposed to a 100 fs, $7 \times 10^{13}$ W/cm$^2$ pulse, using two models 
for screening.  In the first model, the screening length is not 
allowed to fall below xenon's Wigner-Seitz radius at liquid density. 
The second model for screening uses a formula given by Attard.} 
	\label{fig:screeninglengths} 
\end{figure}

As shown in Fig. \ref{fig:screeninglengths}, when plasma screening 
of the Xe ions becomes strong enough to allow  
photoionization of Xe$^{+}$ into Xe$^{2+}$, large numbers of extremely 
low-energy electrons are added to the plasma.  As a result, the ratio 
of electron kinetic energy to electrostatic potential energy falls 
dramatically,  the Debye 
length of the plasma falls abruptly below the Wigner-Seitz radius of xenon, 
and the plasma enters a regime of strong correlation.  In this regime, 
a number of the assumptions of Debye-H\"uckel screening model break 
down, and the Debye length loses its meaning as a screening 
distance \cite{Fortov}.  If the plasma cools sufficiently, screening 
lengths can become complex, and result in oscillatory electron--ion 
correlation functions \cite{lee,attard}.   
 
Another possibly important effect of the strongly coupled plasma was 
identified in a recent 
study \cite{jungreuthmayer05}, which has identified electron dynamics in a 
strongly coupled plasma as having a very large impact upon rates of 
many-body recombination and hence upon energy absorption by the 
cluster as the re-combined ions undergo multiple episodes of 
photoionization. 
 
Most calculations performed in this paper were 
performed using xenon's Wigner-Seitz radius at liquid density 
as a minimum value below which the 
screening was not allowed to fall.  Clearly, with the precise nature 
of screening unknown in the strongly correlated regime, our method of 
calculating atomic properties based on a Debye-screened atomic 
potential acquires a corresponding uncertainty. In an attempt to 
estimate this uncertainty, we have described the evolution of the cluster 
using different models for the screening length in a highly correlated 
plasma. 
 
Our simplest approximation applied xenon's Wigner-Seitz radius at 
liquid density as a 
minimum value below which the screening was not allowed to fall. 
A second model, proposed by Attard \cite{attard}, deals with ions 
having a nonzero radius.  Strictly speaking, the Debye-H\"uckel model 
for plasma screening is invalid except in the limit of ions which have zero 
size.  Attard has shown that in the case where ions 
have a nonzero hard-sphere radius $d$, the screening length  
$\lambda=1/\kappa$ differs from 
the classical Debye-H\"uckel length $\lambda_D=1/\kappa_D$ 
according to  
\begin{equation} 
\kappa=\frac{\kappa_D}{\sqrt{1-(\kappa_{D}d)^{2}/2+(\kappa_{D}d)^{3}/6}} \; . 
\label{attardscreening}
\end{equation} 
This effect becomes important in the domain where $\lambda_D \le d$.

\begin{figure}%%figure updated 
	\begin{center} 
	\subfigure{ 
	\includegraphics[width=3.375in]{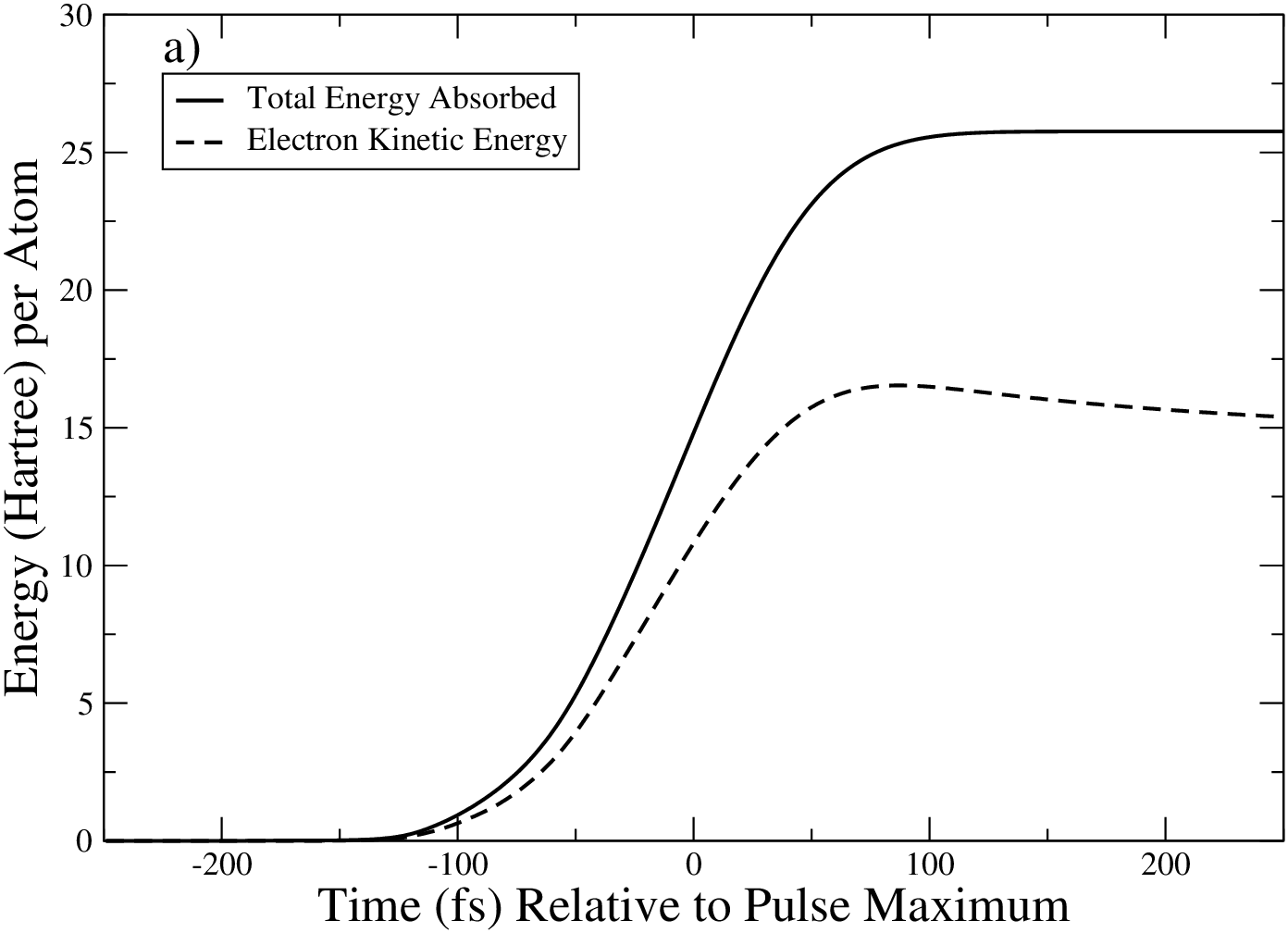}} 
	\vskip 0.15in 
	\subfigure{ 
	\includegraphics[width=3.375in]{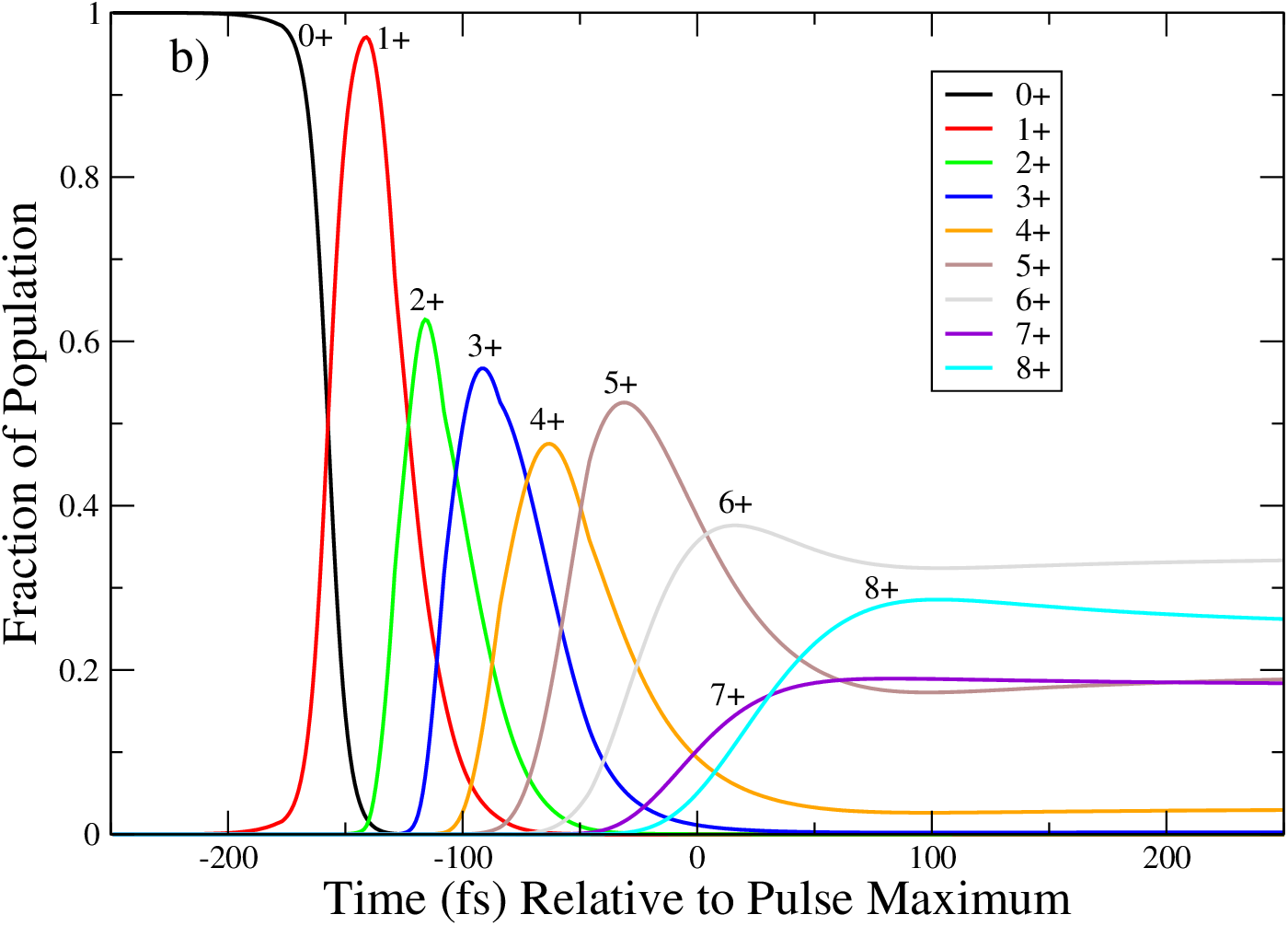}} 
	\end{center} 
	\caption{Near the center of the pulse, the evolution of the
cluster using Attard screening is very similar to the evolution using
Wigner-Seitz screening, shown in Figure \ref{fig:natureparams_debyeWigSeitz}. 
For a 1500 
atom cluster exposed to a $7 \times 10^{13}$ W/cm$^{2}$, 100 fs pulse: 
a) Energy absorbed vs time for the Attard screening model, b) Charge species 
population vs time for the Attard screening model.} 
	\label{fig:ws_vs_attard} 
\end{figure}

\begin{figure}%%figure updated 
	\begin{center} 
	\subfigure{ 
	\includegraphics[width=3.375in]{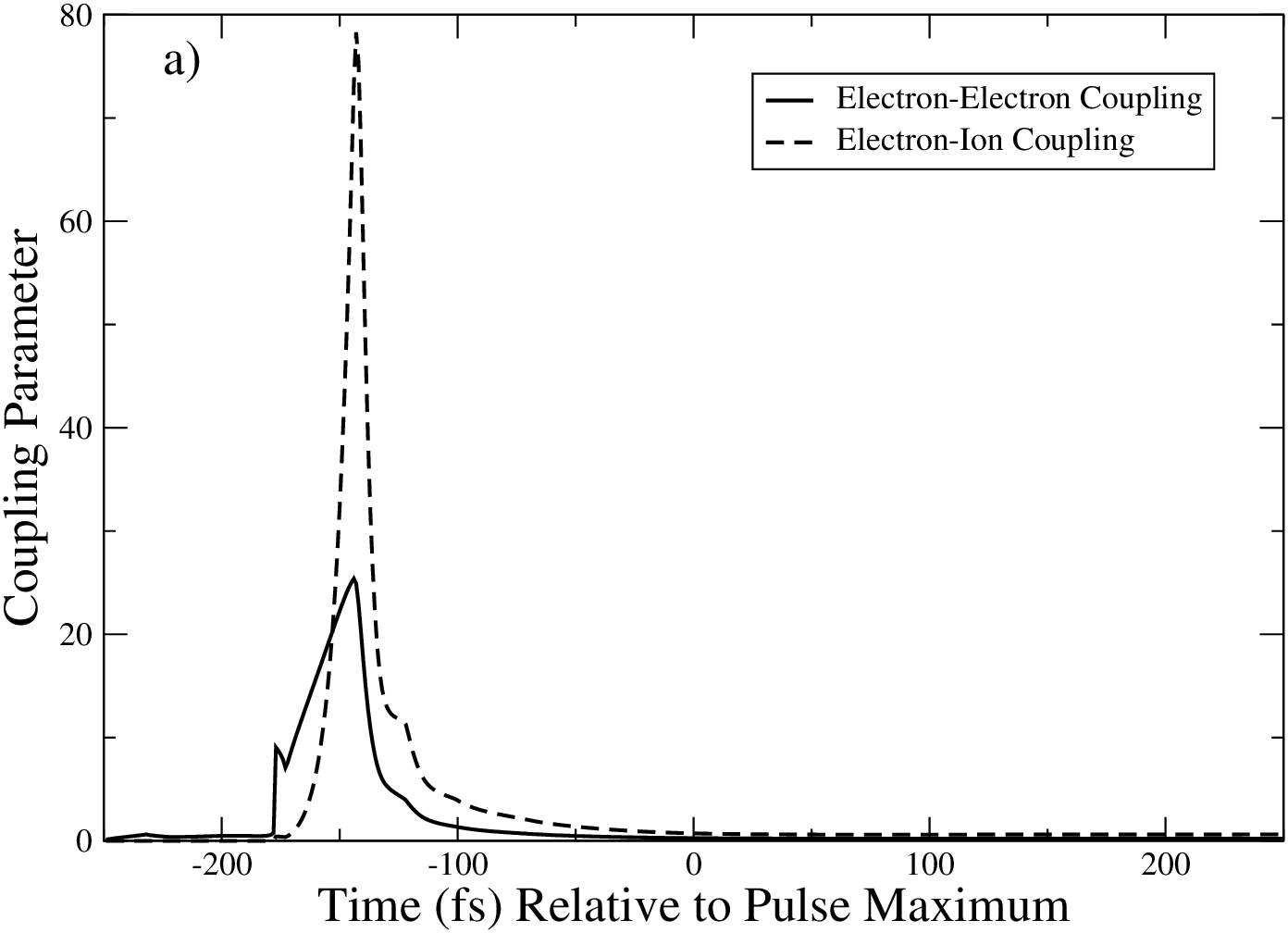}}
	\vskip 0.15in 
	\subfigure{ 
	\includegraphics[width=3.375in]{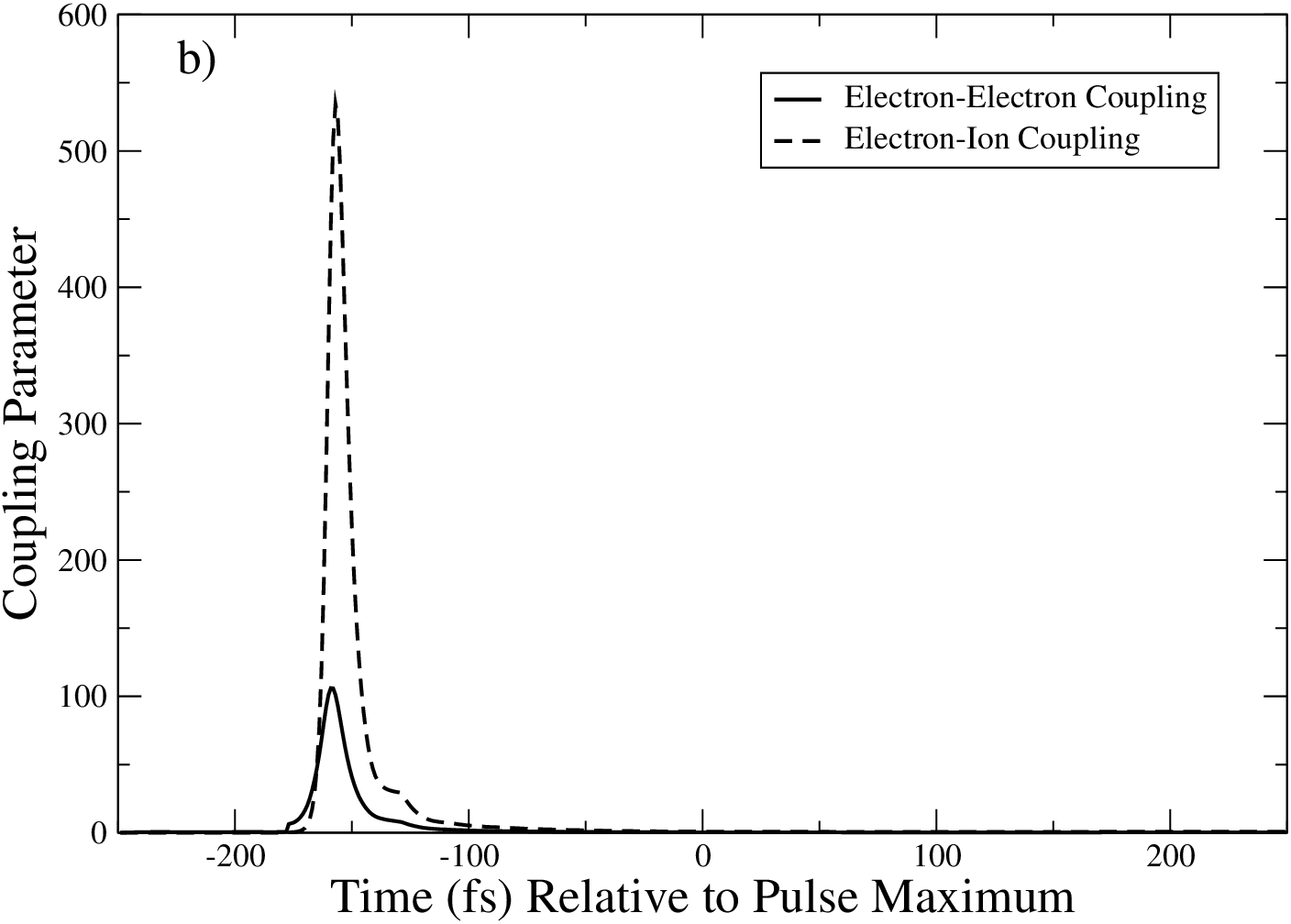}} 
	\end{center} 
	\caption{Plasma coupling parameter vs. time for the two models 
of plasma screening.  The coupling parameters are defined by 
$\Gamma_{ee}=\frac{1}{a T}$ and $\Gamma_{ei}=Z \Gamma_{ee}^{3/2}$
where the average distance between
electrons $a$ is given by $a=(\frac{3}{4 \pi n_{e}})^{1/3}$ and $Z$ is
the average charge of the ions.
The plasma becomes very strongly coupled early in the pulse, but the
strength of the coupling decreases as the plasma absorbs more energy
in the course of the cluster heating. a) Coupling 
parameters vs. time using Wigner-Seitz cutoff. b) Coupling parameters vs. 
time for the Attard screening model.} 
	\label{fig:couplingparameters} 
\end{figure} 
 
Qualitatively, the effect of considering screening lengths in this 
model which are shorter than the Wigner-Seitz radius is twofold. 
First, the tighter screening slightly decreases inverse bremsstrahlung 
heating.  Secondly, it allows 
photoionization of Xe$^{3+}$ and higher charge states.   
Directly substituting the Attard screening length for the Debye length 
with Wigner-Seitz cutoff therefore gives some insight as to how 
sensitive our results are to different models of the ionic potential 
under very strong screening.  As can be seen in Fig. \ref{fig:ws_vs_attard} 
the Attard screening model has a relatively small impact on our 
prediction for the energy absorbed by the cluster. More prominent is 
the formation of higher charge states, which is abetted by the reduced 
ionization potentials resulting from the tighter screening in the 
Attard model.  Figure 
\ref{fig:couplingparameters} shows the plasma coupling parameter, a 
measure for the nonideality of a plasma, for the two models, 
demonstrating that the Attard screening model gives rise to a more 
strongly coupled plasma than the pure Debye model.  In 
addition, the two models give different populations for the various 
charge states at the end of the pulse; however, the combined effects 
of the cluster expansion and spatial averaging over the beam profile 
act to  destroy much of this information. 
 
%\begin{figure}%%new figure 
%	\begin{center} 
%	\subfigure{ 
%	\includegraphics[width=3.375in]{new78pot_chargepop}} 
%	\vskip 0.15in 
%	\subfigure{ 
%	\includegraphics[width=3.375in]{new78pot_chargeenergy}} 
%	\end{center} 
%	\caption{a) Ion population and b) Average kinetic energy 
%vs. charge state for the two sets of input parameters and as recorded 
%in the experiment.  Nature parameters correspond to 1500 atom clusters 
%exposed to a 100 fs, $7 \times 10^{13}$ W/cm$^{2}$ pulse, while Thesis 
%parameters correspond to 2500 atom clusters exposed to a 50 fs, $2.5 
%\times 10^{13}$ W/cm$^{2}$ pulse. 
%The experiment could detect only charged ionic species; 
%accordingly, the theoretical predictions have been normalized such 
%that the sum of the fractional populations for charged species is 1. 
%} 
%	\label{fig:expvstheory} 
%\end{figure} 
 
\section{Hydrogenic Model of Inverse Bremsstrahlung}
\begin{figure} 
	\begin{center} 
	\includegraphics[width=3.375in]{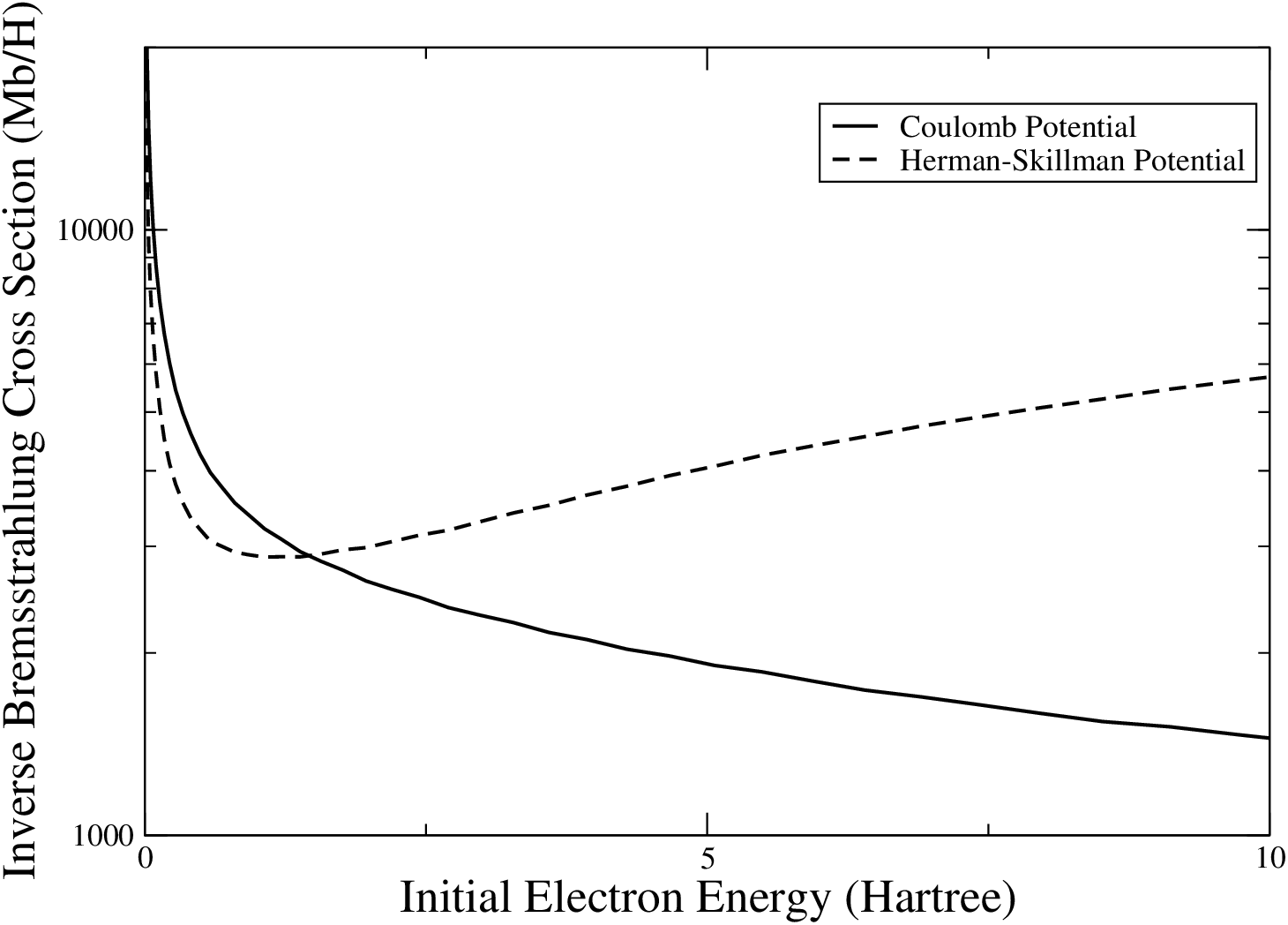} 
	\caption{Inverse bremsstrahlung cross sections
[Eqs. \ref{IBS1} and \ref{IBS2}] calculated for an electron in the
field of a purely Coulombic 5+ potential and for an electron in the
field of a Xe Herman-Skillman atomic potential of the same charge.  In
comparison with figure \ref{fig:IBSComparison}, it can be seen that at
higher charge states, the impact of atomic structure on inverse
bremsstrahlung cross sections is decreased.} 
	\label{fig:5plusIBScomparison} 
	\end{center} 
\end{figure}

Most previous approaches to the problem of laser-cluster interactions
have considered the ionic potential seen by the electron as a pure
Coulomb potential.  This is not an unreasonable approximation: as the
charge of the ion increases, the difference between inverse
bremsstrahlung cross sections calculated using Herman Skillman
potentials and cross sections calculated using Coulomb potentials is
much smaller than in the case of the bare ion.  
This can be seen in figure \ref{fig:5plusIBScomparison}, which
contrasts inverse bremsstrahlung cross sections calculated using
Coulomb and Herman-Skillman potentials for ions of charge 5.

As can be seen in figure \ref{fig:natureparams_debyeWigSeitz}b, when
the laser reaches maximum intensity, most of the cluster has been
ionized to such high charge states.  Thus, models of the inverse
bremsstrahlung process which use Coulombic potentials should be able
to see comparable levels of heating to those using cross sections
derived using Herman-Skillman potentials.

To investigate this proposition, we simulated the laser-cluster
interaction for a 1500 atom cluster exposed to a 100 fs,  
$7 \times 10^{13}$ W/cm$^2$ using our model, but with a physical
picture chosen to emulate that of Siedschlag and Rost
\cite{siedschlag}.  In the
simulation, we used inverse bremsstrahlung cross sections calculated
with Debye-screened Coulomb potentials.  We used the same ionization potentials
and photoionization cross sections as in our other simulations, and
used the unaltered Debye length as the screening length.  Collisional
ionization and recombination were not considered.

\begin{figure}%%figure updated 
	\begin{center} 
	\subfigure{ 
	\includegraphics[width=3.375in]{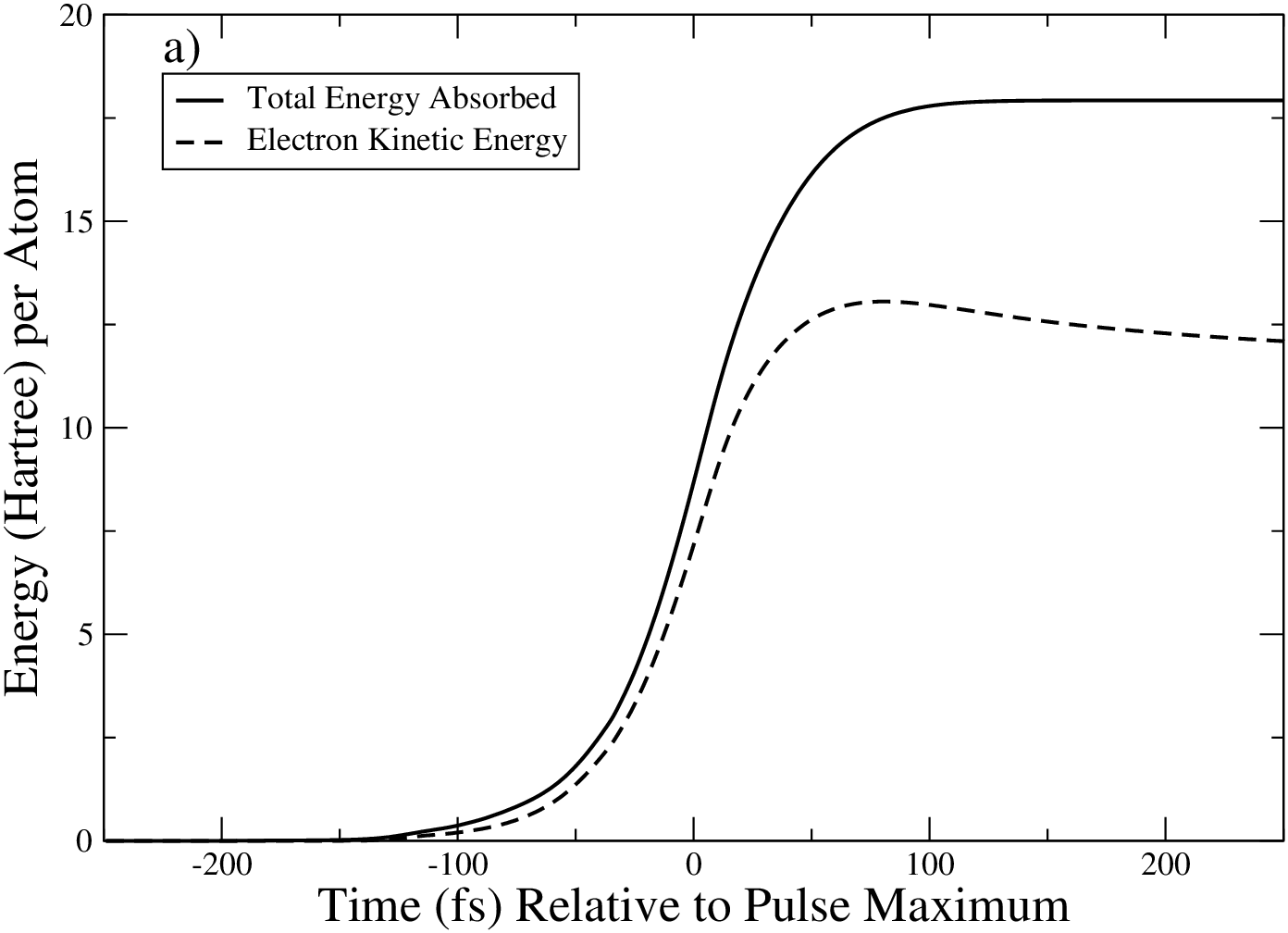}}
	\vskip 0.15in 
	\subfigure{ 
	\includegraphics[width=3.375in]{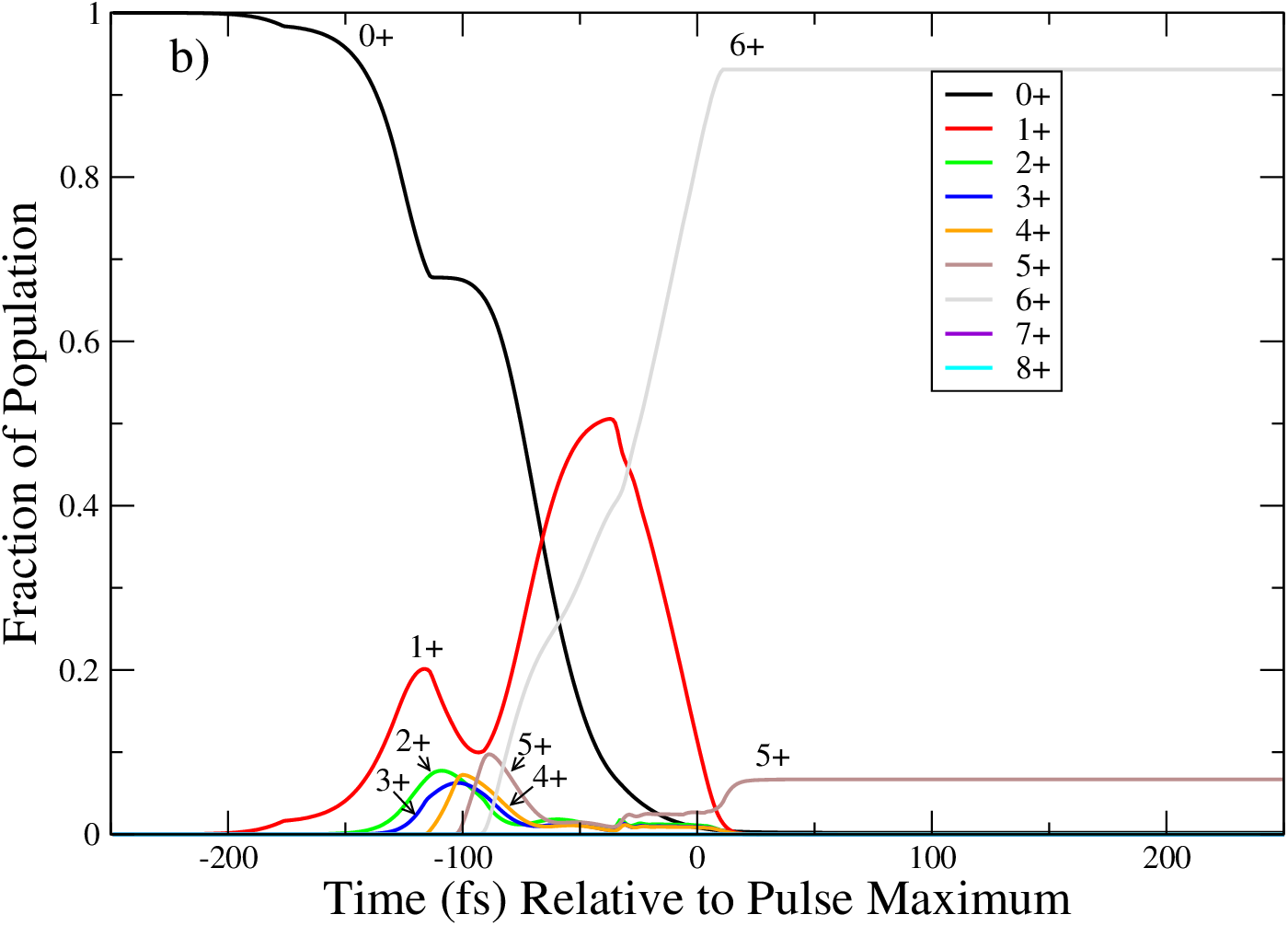}}
	\vskip 0.15in 		
	\subfigure{ 	
	\includegraphics[width=3.375in]{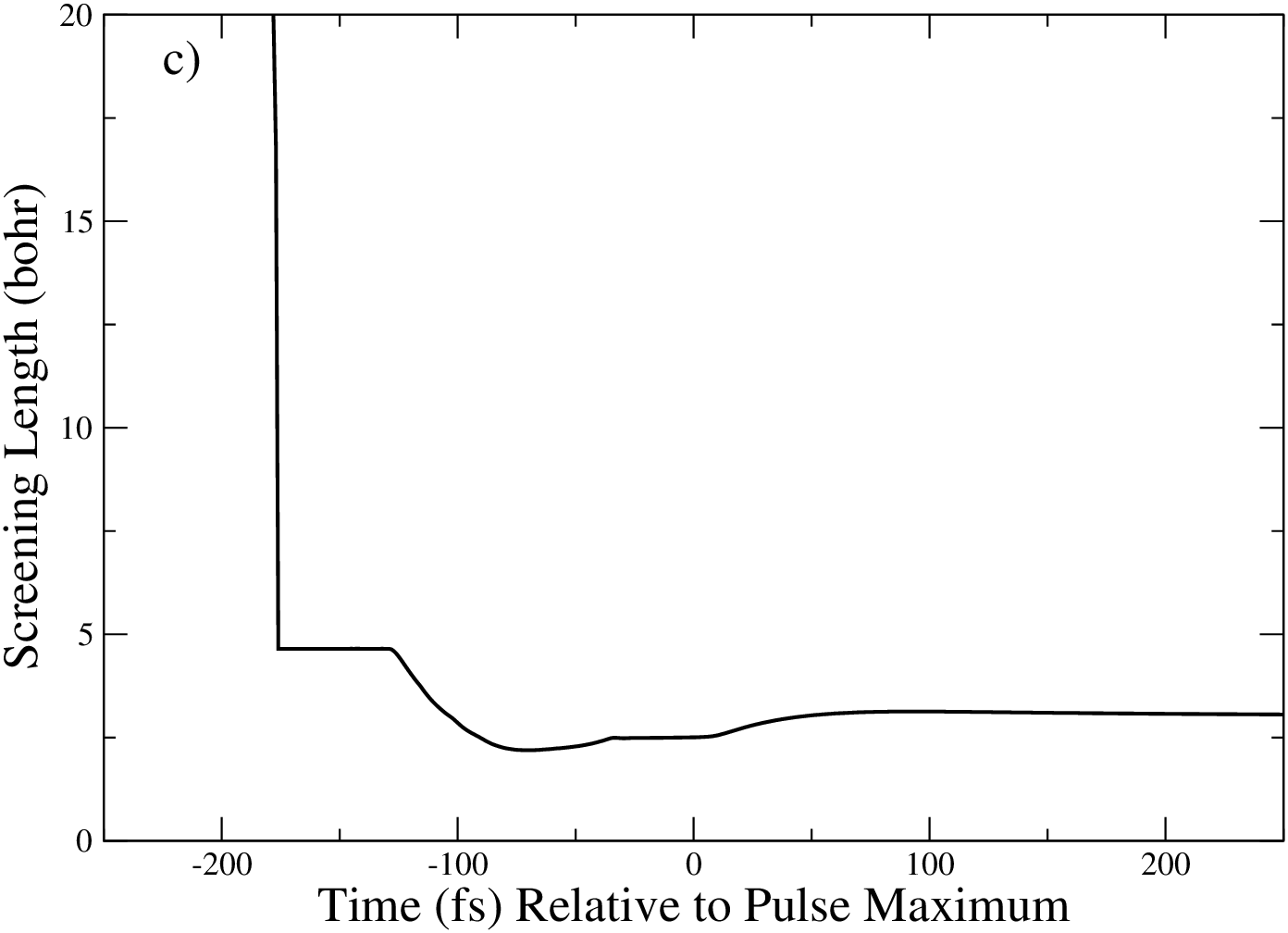}}
	\end{center} 
	\caption{Simulation of the laser-cluster interaction using a
physical model taken from \cite{siedschlag}.  In this model, inverse
bremsstrahlung cross sections are calculated using hydrogenic
potentials and all high charge states are produced via sequential
photoionization.  Collisional ionization and recombination are not
considered.  a)Energy absorbed vs time. b)Charge state population vs
time. c)Debye length vs time.} 
	\label{fig:hydrogenicmodel} 
\end{figure}

The results of this simulation are presented in figure
\ref{fig:hydrogenicmodel}.  We found levels of energy absorption very
comparable to those in our own model but very different behavior of
the ionic populations with time.  Xe$^{7+}$ and
Xe$^{8+}$, which make up almost half of the population of the cluster
at the end of the pulse in our model, were present in negligible
quantities.

Both differences between the two physical pictures are attributable to
the effects of collisional ionization and recombination.
Recombination slows the growth of high charge state populations by
allowing some photoionized ions to recombine into a lower charge
state, while collisional ionization allows the population of charge
states which cannot be created via sequential photoionization.

\section{Conclusions} 
 
When a xenon cluster is irradiated by intense VUV light, there are 
four phases in its evolution.  In the first phase, electrons are 
liberated from the xenon atoms and form a plasma.  As the number of 
free electrons grows, the screening length of the plasma shrinks. 
 
Once the screening length of the plasma reaches 10.6 bohr, 
Xe$^{1+}$ can undergo photoionization 
into Xe$^{2+}$.  This results in the addition of large numbers of low-energy  
electrons to the plasma, cooling it and decreasing the 
screening length still further.  The ratio of kinetic energy to 
potential energy falls dramatically, and the plasma temporarily 
becomes strongly coupled.  Ionization potentials for higher charge 
states fall with increased screening, facilitating their creation. 
 
In the third phase, the plasma undergoes rapid inverse bremsstrahlung 
heating.  High charge states are formed through collisional ionization 
and recombination, and the cluster becomes charged as energetic 
electrons evaporate away from its surface.  The charge state 
distribution shifts rapidly toward higher charges, with the average 
ionic charge reaching 5.5 at the pulse peak.  This distribution 
changes only slowly  on the timescale of the pulse. 
 
%Our improved calculations of inverse bremsstrahlung cross sections 
%are smaller than those found in \cite{santra} by as much as an order 
%of magnitude.  In some sense, then, the results cited in \cite{santra} 
%were fortuitous.  

Finally, the cluster expands due to the pressure of the electron gas 
and the cluster's own charge.  As the cluster expands, the electron 
plasma cools and becomes more diffuse.  Screening lengths increase, 
and charge state equilibrium shifts toward lower charge states.   
 
Of these four phases, our current model describes the first and third 
phases well; the second more crudely. The dynamics of the expanding
cluster are a challenging problem in their own right, and demand a
treatment more sophisticated than our simple homogeneous expansion model.
 
For strongly coupled plasmas, it is unclear whether our treatment of 
plasma screening adequately describes the potential seen by scattering 
or photoionizing electrons.  As the Debye length falls below the 
Wigner-Seitz radius, the interaction of screening effects due to 
inner-shell electrons and effects due to screening by continuum plasma 
electrons should be considered.  It is known that the screening length 
diverges from the Debye length in this limit, but the precise 
nature of the electron--ion potential is unknown.   
 
There is some difficulty in comparing our results to the Hamburg 
experiment, due to experimental uncertainty in laser intensity, 
temporal profile, spatial profile, and cluster size.   
Whereas in the Nature paper the 
Hamburg group described the laser pulse as 100 fs,  
$7 \times 10^{13}$ W/cm$^2$ incident on 1500 atom xenon clusters, 
Wabnitz's thesis \cite{Wabnitz03} subsequently describes these pulses as  
50 fs, $2.5 \times10^{13}$ W/cm$^2$ pulses incident on 2500 atom 
clusters.  In addition, the temporal profile of the laser pulses is 
not Gaussian, and varies in an unpredictable way from pulse to pulse 
due to the nature of the SASE amplification process, which starts from 
shot noise. 

Our model also has difficulty explaining the properties of the
clusters long after the laser-cluster interaction is over.  As the
clusters expand and cool, they continue to undergo collisional
ionization and recombination.  The distribution of charge states
measured at the experimental detectors bears no simple relationship to
the distribution we calculate at the end of the pulse.  Our
homogeneous model of the cluster expansion implicitly requires that
all charge states in the same cluster have the same average kinetic
energy; this obviously conflicts with the quadratic dependence of
energy vs charge state detected in the Hamburg experiment.  Also, it
is likely that high charge states escape the cluster more quickly than
low charge states, spending less time in regions of high electron
density and having less opportunity to recombine.  Thus, a
more sophisticated model of the cluster expansion is necessary in
order to predict final charge state and ionic energy distributions
with confidence for
comparison with experiment. 
 
At the center of a gaussian laser pulse using parameters taken from
the Nature paper and a Wigner-Seitz debye length cutoff, each cluster
absorbs on average 682 eV per atom.  At a distance of 3 sigma from the
center of such a gaussian pulse, each cluster absorbs only .4 eV per
atom.  Spatial averaging over the gaussian pulse profile from 0 to 3
sigma gives an average of 195 eV per atom absorbed.  Using parameters
taken from Wabnitz's thesis gives 219 eV per atom at the center, 0.2 eV
per atom at 3 sigma, and 65 eV per atom on spatial averaging.

Using a time of flight detector which could detect only charged ions, Wabnitz 
{\em et al.} reported an average ion energy of 400 eV, subsequently revised 
to 650 eV.

Clearly, a spatial average such as we perform could be altered by
averaging over a different beam profile or by changing the limits of
the radial average and including more clusters which are exposed to
only a tiny fraction of the beam's peak intensity.  It is also clear
that most of the atoms in the clusters which are exposed to very small
fractions of the peak intensity will never be ionized and thus would
not register in a time-of-flight ion detector such as was used in the
Hamburg experiment.  Thus, in the absence of better information about
the beam's spatial and temporal profile and a more comprehensive model
of the cluster expansion after the conclusion of the laser pulse, it
is impossible to make precise comparisons between our model and the
Hamburg results. 

Nevertheless, our model of the laser-cluster interaction explains some
surprising features of the laser-cluster interaction in the VUV regime
quite well.  Primary among these is the surprising efficiency by which
the clusters absorb photons.  Second, we explain the origin of the
high charge states observed in the Hamburg experiment.
Third, we have with the same model calculated the early electron 
ejection spectrum measured in \cite{laarman05} and achieved great 
similarity to experiment, despite a cluster size and pulse intensity 
which differ significantly from those of the original Hamburg 
experiment.  We have shown that such spectra can depend 
strongly on the model of plasma screening or the precise parameters of 
the experiment, and can therefore serve as a possible window into the 
nature of the laser-cluster dynamics during the time period of the pulse. 
 
In conclusion, we have introduced a model of the laser--cluster 
interaction in the VUV regime which takes into account improved 
calculations of inverse bremsstrahlung heating, photoionization, 
collisional ionization and recombination.  The effects of plasma 
screening on all of these processes are included, and an alternative 
model of very strong plasma screening has been considered. 
  
\acknowledgments 
We would like to thank Thomas M\"oller, Hubertus Wabnitz and Tim
Laarmann for valuable 
information about their experiments and Jan-Michael Rost for
stimulating discussions.   This work was supported in part by 
the Office of Basic Energy Sciences, Office of Science, U.S. Department 
of Energy; R.S. under Contract No. W-31-109-ENG-38. 
 
%\bibliography{references} 

\begin{thebibliography}{47}
\bibitem[{com()}]{comment1}
\bibinfo{note}{In practice, matrix elements involving large values of $l$ were
  found to converge very quickly to zero. Because of this, the total transition
  rate from states of energy $E$ to states of energy $E^{\prime}$, and hence
  the heating, is unaffected by the existence of a cutoff.}
\end{thebibliography}

\end{document}